\documentclass[twocolumn,superscriptaddress,nopacs,aps,prl]{revtex4-2}

\usepackage{graphicx}  
\usepackage{dcolumn} 
\usepackage{bm}  
\usepackage{amsmath}
\usepackage{amsthm}
\usepackage{amssymb}

\usepackage[T1]{fontenc}
\usepackage[utf8]{inputenc}
\usepackage[normalem]{ulem}

\usepackage{hyperref}

\usepackage{inputenc}
\usepackage{xcolor,multirow}
\usepackage{cleveref, url}
\usepackage{amsfonts}
\usepackage{bbold}
\usepackage{mathrsfs}
\usepackage{mathtools}
\usepackage{pgfplots}
\usepackage{physics}

\usepackage{cancel}
\usepackage{tcolorbox}
\usepackage{float}
\usepackage{upgreek}
\usepackage{tikz}

\newcommand{\be}{\begin{equation}}
\newcommand{\ee}{\end{equation}}

\NewDocumentCommand{\scinot}{m m o}{%
    \ensuremath{%
        #1 \times 10^{#2}%
        \IfValueT{#3}{\,\mathrm{#3}}%
    }%
}

\begin{document}

\title{Enantioselective optical trapping and characterization of all dielectric disorder-enabled chiral particles }

\author{Guilherme T. Moura}
 \email{guilhermeluis@pos.if.ufrj.br}
\affiliation{Instituto de F\'{\i}sica, Universidade Federal do Rio de Janeiro \\ Caixa Postal 68528,   Rio de Janeiro,  Rio de Janeiro, 21941-585, Brazil}
\affiliation{CENABIO - Centro Nacional de Biologia Estrutural e Bioimagem \\
Universidade Federal do Rio de Janeiro, Rio de Janeiro, Rio de Janeiro 21941-902, Brazil}
\author{Tanja Schoger}
\email{tanja.schoger@lkb.upmc.fr }
 \affiliation{Laboratoire Kastler Brossel, Sorbonne Université, CNRS, ENS-PSL, Collège de France, 75005 Paris, France}
 \author{Kainã G. Diniz}
\affiliation{Instituto de F\'{\i}sica, Universidade Federal do Rio de Janeiro \\ Caixa Postal 68528,   Rio de Janeiro,  Rio de Janeiro, 21941-585, Brazil}
\affiliation{CENABIO - Centro Nacional de Biologia Estrutural e Bioimagem \\
Universidade Federal do Rio de Janeiro, Rio de Janeiro, Rio de Janeiro 21941-902, Brazil}
\author{Marcel A. B. Morte}
\affiliation{Instituto de F\'{\i}sica, Universidade Federal do Rio de Janeiro \\ Caixa Postal 68528,   Rio de Janeiro,  Rio de Janeiro, 21941-585, Brazil}
\affiliation{CENABIO - Centro Nacional de Biologia Estrutural e Bioimagem \\
Universidade Federal do Rio de Janeiro, Rio de Janeiro, Rio de Janeiro 21941-902, Brazil}
\author{Diney S. Ether Jr}
\affiliation{Instituto de F\'{\i}sica, Universidade Federal do Rio de Janeiro \\ Caixa Postal 68528,   Rio de Janeiro,  Rio de Janeiro, 21941-585, Brazil}
\affiliation{CENABIO - Centro Nacional de Biologia Estrutural e Bioimagem \\
Universidade Federal do Rio de Janeiro, Rio de Janeiro, Rio de Janeiro 21941-902, Brazil}
\author{Leonardo de S. Menezes}
\affiliation{Chair in Hybrid Nanosystems, Nanoinstitute Munich, Faculty of Physics, Ludwig-Maximilians-University, 80539 Munich, Germany
}
\affiliation{
Departamento de Física, Universidade Federal de Pernambuco, 50670-901 Recife-PE, Brazil}
\author{Yicui Kang}
\affiliation{Chair in Hybrid Nanosystems, Nanoinstitute Munich, Faculty of Physics, Ludwig-Maximilians-University, 80539 Munich, Germany
}
\author{Emiliano Cort\'es}
\affiliation{Chair in Hybrid Nanosystems, Nanoinstitute Munich, Faculty of Physics, Ludwig-Maximilians-University, 80539 Munich, Germany
}
\author{Cyriaque Genet}
 \affiliation{Université de Strasbourg, CNRS, Institut de Science et d’Ingénierie Supramoléculaires, UMR 7006, F-67000 Strasbourg, France}
\author{Nathan B. Viana}
\email{nathan@if.ufrj.br}
\affiliation{Instituto de F\'{\i}sica, Universidade Federal do Rio de Janeiro \\ Caixa Postal 68528,   Rio de Janeiro,  Rio de Janeiro, 21941-585, Brazil}
\affiliation{CENABIO - Centro Nacional de Biologia Estrutural e Bioimagem \\
Universidade Federal do Rio de Janeiro, Rio de Janeiro, Rio de Janeiro 21941-902, Brazil}
\author{Felipe A. Pinheiro}
 \email{fpinheiro@if.ufrj.br }
\affiliation{Instituto de F\'{\i}sica, Universidade Federal do Rio de Janeiro \\ Caixa Postal 68528,   Rio de Janeiro,  Rio de Janeiro, 21941-585, Brazil}
\author{Paulo A. Maia Neto}
 \email{pamn@if.ufrj.br}
\affiliation{Instituto de F\'{\i}sica, Universidade Federal do Rio de Janeiro \\ Caixa Postal 68528,   Rio de Janeiro,  Rio de Janeiro, 21941-585, Brazil}
\affiliation{CENABIO - Centro Nacional de Biologia Estrutural e Bioimagem \\
Universidade Federal do Rio de Janeiro, Rio de Janeiro, Rio de Janeiro 21941-902, Brazil}%

\date{\today}% 

\begin{abstract}
We trap submicroscopic silica spheres coated with randomly distributed titanium dioxide nanoparticles in optical tweezers with Laguerre–Gaussian modes and observe orbital dynamics that differ from those of achiral silica spheres. 
We show that the disordered nanoparticle coating generates an effective chiral geometry, giving rise to enhanced enantioselective chiral optical forces and 
a measurable modification of the orbital period. 
A theoretical model based on the Mie–Debye formalism, including optical aberrations, not only quantitatively explains the experimental results but also allows to characterize the Pasteur parameter quantifying the chiroptical response of individual composite particles. 
These findings constitute direct experimental evidence of chiral optical forces exerted by structured light beams on individual chiral particles and identify disorder-enabled, all-dielectric particles as a versatile material platform to tailor chiral optical forces at the nanoscale.
\end{abstract}

\maketitle

\textit{Introduction---}Chirality, the lack of superposition between an object and its mirror image, is ubiquitous in nature, from molecules to galaxies~\cite{wagniere2007chirality}. 
In optics, it gives rise to circular dichroism and optical rotation through differential interactions with opposite circular polarizations~\cite{wagniere2007chirality}. Distinguishing enantiomers is crucial in chemistry, biology, and pharmacology, where handedness governs biological activity and drug efficacy~\cite{sui2023strategies,scriba2002selected}. 
However, conventional enantioseparation methods are often invasive and system-specific~\cite{gubitz2001chiral}, motivating the development of optical approaches for chiral discrimination.

Conventional chiroptical techniques are fundamentally limited by the weak chiral response of matter~\cite{lindell1994electromagnetic}. 
Although plasmonic and metamaterial platforms can substantially enhance chiral light--matter interactions~\cite{Ogier2014,mohammadi2023nanophotonic,mohammadi2021dual,solomon2020nanophotonic,kim2022enantioselective,loren2023circular,schnoering2018three,gryb2023two,kuhner2023unlocking,Olmos-Trigo2024,heimig2026chiral,cerdan2023chiral}, they predominantly probe ensemble-averaged responses. 
Consequently, enantioselective manipulation and characterization of individual particles \cite{Olmos-Trigo2026, dutra2026circular} and molecules remains challenging because the Pasteur parameter $\kappa$, which quantifies the chiral electromagnetic response, is typically very small~\cite{mohammadi2023nanophotonic}.

Chiral optical forces provide a promising route for the manipulation, transport, and mechanical separation of individual enantiomers~\cite{genet2022chiral,Nori2015_6,Bliokh3300_7,Zhao1055_8,Schnoering0410_9,Bliokh0336_11,wang2014lateral,zhang2017all,lai2024observation,feng2026revealing,cao2019fano,yamanishi2022optical,zhang2024enantioselective,man2024construction,ayuso2021ultrafast,li2022enantioselective,schied2023chirality,wen2025optical,fang2021optical,yao2024sorting,xu2025optical,li2021optical,li2026enantioselective}. 
Existing theoretical~\cite{canaguier2013mechanical,Hayat1319_13,Genet0438_14,Zhang4292_15,Cao1801_16,Zheng4515_17,Zhang6282_18,Chan3307_19,Ho2017_20,Solomon2019_21,hentschel2017chiral,lin2021plasmonic,zhao2016enantioselective,martinez2024longitudinal} and experimental~\cite{Brasselet2018_10,tkachenko2014optofluidic,Shi2020_24,Tkachenko2014} approaches mainly follow two strategies. 
The first exploits engineered material platforms, particularly plasmonic nanostructures~\cite{hentschel2017chiral,lin2021plasmonic,zhao2016enantioselective}, to enhance chiral optical forces. 
However, plasmonic systems suffer from intrinsic losses and thermoplasmonic effects~\cite{govorov2007generating}. 
This has motivated low-loss dielectric alternatives, including integrated waveguides~\cite{martinez2024chiral} and optical fibers~\cite{golat2024optical}, where evanescent fields enable enantiomeric particle sorting. 
The second strategy employs free-space illumination, as in optical tweezers~\cite{Ali2020,ali2021enantioselection,ali2020probing,ali2023enantioselective,Diniz2025}, or structured light to achieve all-optical chiral discrimination~\cite{brullot2016resolving,kerber2018orbital,bliokh2015spin,li2021optical}. 
Despite these advances, most enantioselective schemes rely on strong field gradients that enhance both chiral and achiral gradient forces. 
Consequently, the achiral contribution generally dominates unless the particle exhibits very large values of $\kappa$, limiting trapping efficiency and applicability to naturally occurring chiral particles and molecules.

\begin{figure*}[htb]
\centering

\begin{tikzpicture}
\node[inner sep=0] (img)
  {\includegraphics[width=0.5\linewidth]{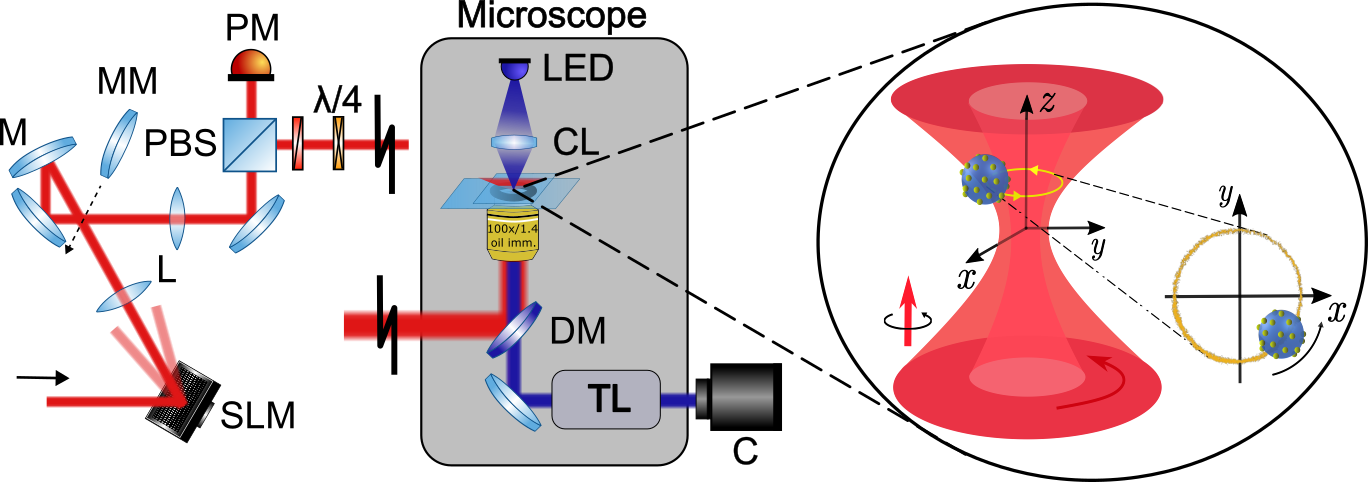}};
\node[anchor=north west, font=\scriptsize] 
  at ([xshift=-15pt,yshift=0pt] img.north west) {(a)};
  \node[anchor=north west, font=\scriptsize] 
  at ([xshift=152pt,yshift=0pt] img.north west) {(b)};
\end{tikzpicture}
\raisebox{-0.49cm}{
\begin{tikzpicture}
\node[inner sep=0] (img)
  {\includegraphics[width=0.25\linewidth
  ]{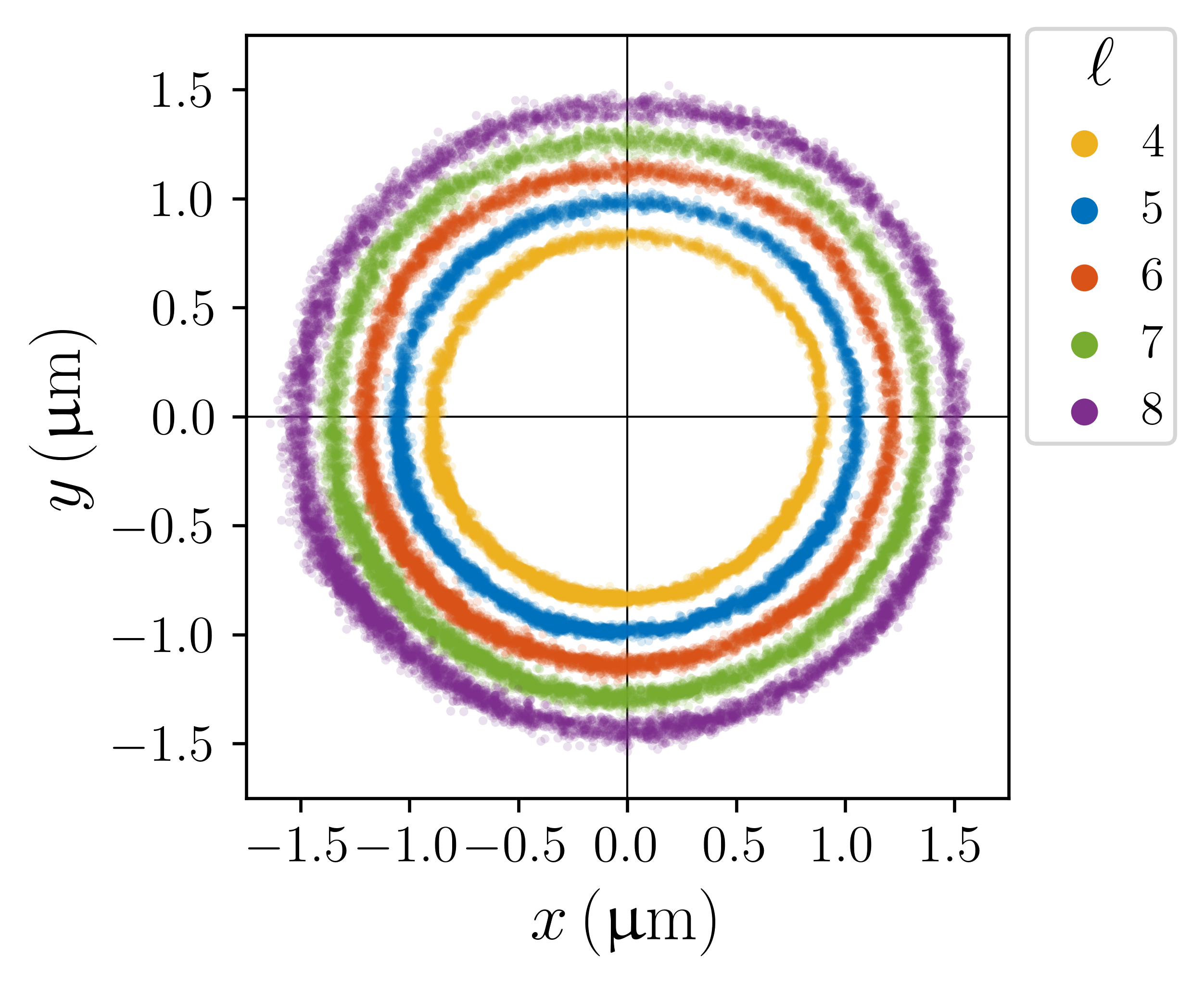}};
\node[anchor=north west, font=\scriptsize] 
  at ([xshift=-3pt,yshift=-1pt] img.north west) {(c)};
\end{tikzpicture}
}
\hspace{-11pt}
\raisebox{0cm}{
\begin{tikzpicture}
\node[inner sep=0] (img)
  {\includegraphics[width=0.19\linewidth]{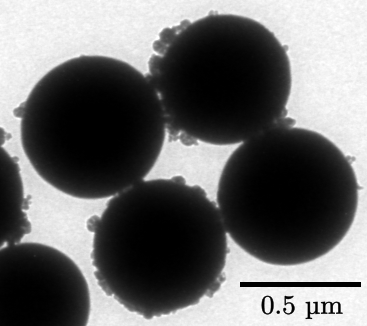}};
\node[anchor=north west, font=\scriptsize] 
  at ([xshift=-3pt,yshift=3pt] img.north west) {(d)};
\end{tikzpicture}
}
\caption{(a) Experimental setup: L: converging lens, M: dielectric mirror, MM: movable mirror, SLM: spatial light modulator,  $\lambda / 4$: quarter-wave plate, PBS: polarizing beam splitter, PM: power meter, DM: dichroic mirror, CL: condenser lens, LED: light-emitting diode, TL: tube lens, C: camera. (b) Schematic representation of a composite bead performing an orbital motion while trapped by a LG mode. (c) Experimental orbit plot of a single composite bead trapped by LG modes $\ell = 4,..., 8$ with left-circular polarization. (d) Transmission electron microscopy images of composite beads, showing nanoparticles of $\text{TiO}_2$ adhered to the surface of silica submicrometric spheres. } 
\label{fig:setup}
\end{figure*}
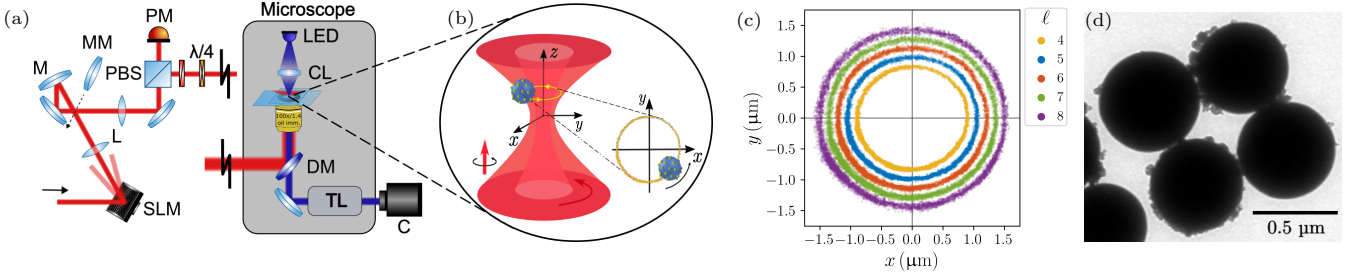

Here we demonstrate enantioselective optical trapping by suppressing the achiral component of the optical force in the measurement strategy, enabling direct observation of chiral forces on individual beads. 
To enhance the chiroptical response, we use circularly-polarized structured Laguerre–Gaussian (LG) beams and measure the resulting orbital motion of all-dielectric particles with geometrical chirality induced by engineered disordered coatings, as illustrated in Fig.~\ref{fig:setup}. 
We isolate the chiral optical force component by comparing the orbital motions induced by beams with opposite handedness. 
We find Pasteur parameters as large as $\kappa\sim 10^{-2}$. A Mie–Debye model~\cite{Diniz2025} quantitatively explains the measurements, establishing disorder-enabled chiral structures~\cite{pinheiro2017spontaneous} as a versatile platform for tailoring chiral optical forces on individual submicroscopic particles.

\textit{Experimental methods---}As shown in Fig.~\ref{fig:setup}(a), a horizontally polarized TEM$_{00}$ laser beam (IPG Photonics, $\lambda=1064\,$nm) emerges from an optical fiber. 
Then, it passes through a telescope expanding its transverse beam size by a factor of 1.5, before illuminating a spatial light modulator (SLM, Holoeye Photonics). 

We structure both amplitude and phase to generate Laguerre-Gaussian modes $\text{LG}_p^{|\ell|}$ with fixed radial number $p = 0$ and topological charges $\ell = \pm 4, \pm 5, \pm 6, \pm 7, \pm 8$. The structured light field is relayed by a $4f$ system onto the objective's back focal plane.
To monitor the laser power alongside the orbital motion of the composite particles, a fraction of the beam is diverted to a power meter.

We combine spatial and polarization degrees of freedom to prepare spin-orbit modes of opposite handedness, ensuring that the topological charge and helicity have the same sign in all cases. 
The sign of the former is switched by using a movable mirror, as reflection acts as a parity transformation on the beam's wavefront. 
For the latter, left- and right-circular polarizations (LCP and RCP) are prepared using a combination of half- and quarter-wave plates to precompensate polarimetric aberrations introduced by the dichroic mirror, which reflects the laser wavelength and transmits the LED light (470\,nm)  illuminating the sample. 
Using laser beams with opposite spin-orbit handednesses 
allows one to isolate the chiral force contributions and cancel out the achiral ones.

The laser beam is then focused by an objective (100x, oil immersion, NA = 1.4, entrance radius $R_{\text{obj}} = 2.8\,$mm) into a sample chamber, formed by two glass coverslips joined by an o-ring, in which the particles are suspended in water. 
The particle image is projected onto a CMOS camera (Hamamatsu Orca-Flash 2.8), operating at an acquisition rate of 100 frames per second. 
The tightly-focused vortex beam traps a single dielectric particle while driving it along an orbital motion as illustrated in Fig.~\ref{fig:setup}(b). 
In order to achieve orbits closer to the circular geometry with approximately uniform velocities, the SLM is employed not only to structure the LG mode but also to correct most optical aberrations. 
We use the information from measurements of the orbit shape and position distribution along the trajectory to determine the optimal phase mask in the SLM  as in Ref.~\cite{Otsu-Hyodo2024}. The resulting experimental orbits for the composite particles are shown in Fig.~\ref{fig:setup}(c).

We optically trap spherical silica particles (control samples) and silica particles decorated with TiO$_2$ nanoparticles. Representative transmission electron microscopy images of the composite particles are shown in Fig.~\ref{fig:setup}(d), while synthesis details are provided in the Supplemental Material (SM)~\cite{supplement}. 
The random arrangement of TiO$_2$ nanoparticles is at the origin of the geometric chirality and hence the associated chiroptical response~\cite{pinheiro2017spontaneous,ali2020probing}. 
Since the constituent materials are intrinsically achiral, the resulting chirality is purely geometric~\cite{rebholz2024separating}, establishing an all-dielectric, disorder-enabled platform for assembling submicrometer chiral particles. 
 
\textit{Theoretical Analysis---}The chiroptical properties of the composite particles are theoretically modeled as a core-shell geometry using an effective-medium approach~\cite{Pastoriza-Santos2004}, as described in SM~\cite{supplement}. 
The silica core is modeled as an achiral homogeneous sphere, while the arrangement of TiO$_2$ nanoparticles is modeled as a chiral shell using the Maxwell-Garnett theory parametrized by a filling fraction and an effective Pasteur parameter $\kappa$. 

The period of a particle moving in an optical force field on a closed orbit is given by
\begin{equation}\label{eq:T}
    T = \beta \int_0^{2\pi}  \frac{\rho_\text{eq}(\phi)}{\vert F_\phi(\rho_\text{eq}, \phi, z_\text{eq})\vert}d\phi\,,
\end{equation}
where the parameter $\beta$ defines the viscous drag experienced by a spherical particle moving in a fluid near a planar boundary, with the Faxén correction \cite{Goldman1967, Schaffer2007} taken into account (see SM~\cite{supplement} for details). 
The theoretical analysis centers on the azimuthal optical force component $F_\phi$, which drives the particle along its stable orbit, parameterized by $(\rho_\mathrm{eq}(\phi), z_\mathrm{eq}(\phi))$ in cylindrical coordinates. 
Indeed, $F_\phi$ captures most of the composite particle's chiroptical response, whereas the trajectory 
is roughly independent of the laser beam spin-orbit handedness~\cite{Diniz2025}. Nevertheless, we also account for the small effect on $\rho_\mathrm{eq}(\phi)$ caused by the relative handedness between the particle and the field.

\begin{figure}
\centering
    \begin{minipage}[b]{1\linewidth}
\centering
\begin{tikzpicture}
\node[inner sep=0] (img)
  {\includegraphics[width=1\linewidth]{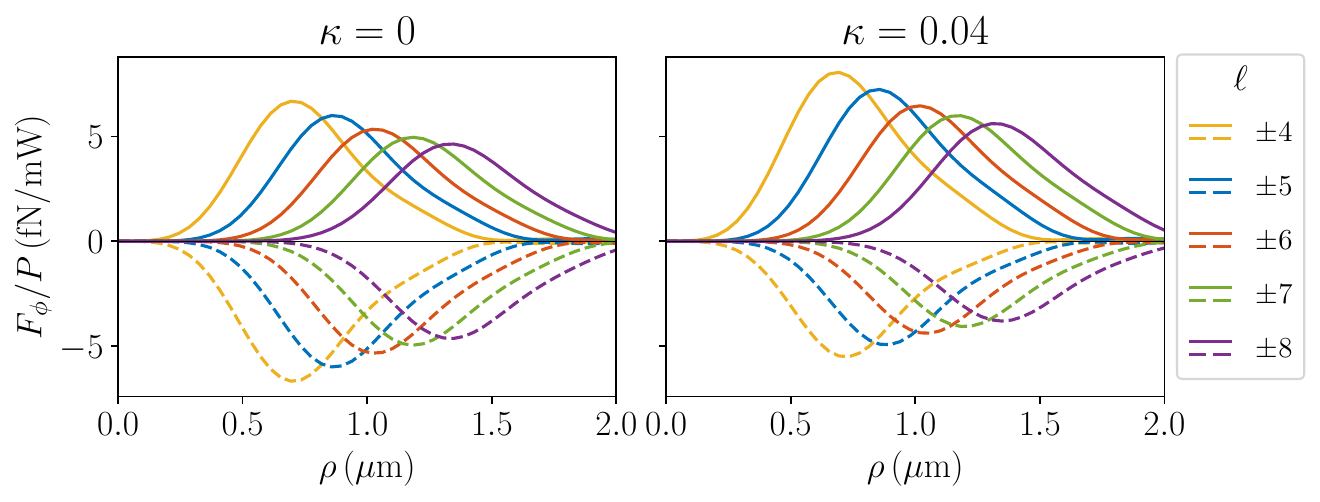}};
\node[
  anchor=north west,
  font=\scriptsize
] at ([xshift=10pt,yshift=2pt] img.north west) {(a)};
\node[
  anchor=north west,
  font=\scriptsize
] at ([xshift=4cm,yshift=2pt] img.north west) {(b)};
\end{tikzpicture}
\end{minipage}

    \begin{minipage}{.45\linewidth}
\centering
\begin{tikzpicture}
\node[inner sep=0] (img)
  {\includegraphics[width=1\linewidth]{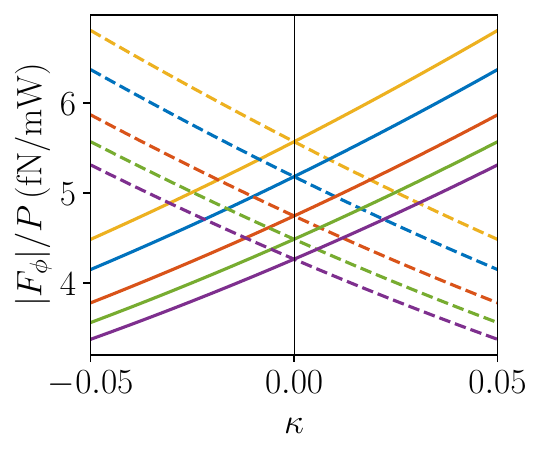}};
\node[
  anchor=north west,
  font=\scriptsize
] at ([xshift=0pt,yshift=6pt] img.north west) {(c)};
\end{tikzpicture}
\end{minipage}
\hspace{0em}
\begin{minipage}{.45\linewidth}
\centering
\begin{tikzpicture}
\node[inner sep=0] (img)
  {\includegraphics[width=1\linewidth]{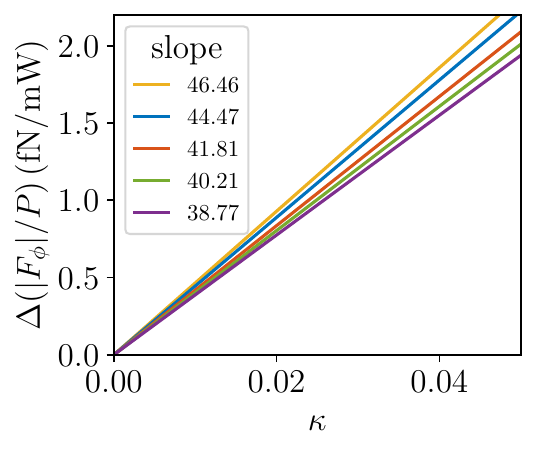}};
\node[
  anchor=north west,
  font=\scriptsize
] at ([xshift=0pt,yshift=6pt] img.north west) {(d)};
\end{tikzpicture}
\end{minipage}

\caption{Azimuthal optical force divided by the optical power $F_\phi/P$ as a function of the radial particle displacement $\rho$ computed at the  equilibrium plane $z=z_\text{eq}$
for (a) an achiral and (b) a chiral core-shell particle characterized by the Pasteur parameter $\kappa=0.04$ and filling fraction $f=0.1$. 
(c) Absolute value of the azimuthal force (in units of power), computed along the stable orbit $(\rho_\text{eq}, z_\text{eq}),$  as a function of the Pasteur parameter $\kappa$. 
(d) Force difference $\Delta (|F_\phi|/P)$ representing the chiral optical force as a function of $\kappa$. The inset shows the results for the slope (in fN/mW) as derived from a linear fit. }
\label{fig:f_phi}
\end{figure}
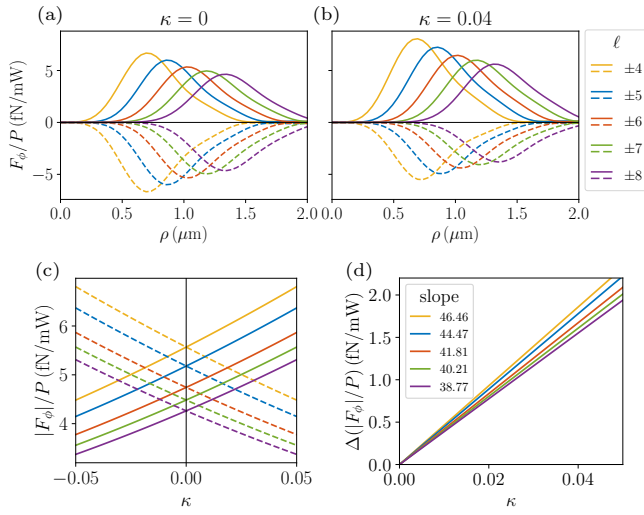

We model the optical force by combining a Debye-like nonparaxial model for the focused Laguerre-Gaussian beam~\cite{Monteiro2009,Fonseca2024} with Mie scattering by spherical core-shell particles~\cite{Ali2020}.  
For achiral particles, parity symmetry imposes that $F_{\phi}\to -F_{\phi}$ when the positive ($|\ell|$, LCP) and negative ($-|\ell|$, RCP) laser beams are interchanged.
On the other hand, the two beams with opposite handedness interact differently with chiral particles, thus leading to distinct values of $|F_{\phi}|.$ 
Those properties are demonstrated in Figs.~\ref{fig:f_phi}(a) and \ref{fig:f_phi}(b), which show the variation of $F_{\phi}/P$ ($P=$ power in the sample region) with the radial component of the particle position, for both achiral and chiral core-shell spheres, respectively. 
In both cases, the annular region of maximum energy density increases linearly with $|\ell|$ \cite{Curtis2003,Monteiro2009}, thus displacing the peak of $F_{\phi}$ away from the axis, while the peak values decrease with $|\ell|.$ 
Figure~\ref{fig:f_phi}(c) shows that the azimuthal force is enhanced when the chiral particle and the laser beam share the same handedness. 
In addition, the difference $\Delta \vert F_\phi\vert =|F_{\phi,+|\ell|}|-|F_{\phi,-|\ell|}|$ represents twice the chiral component of the optical force. 
$\Delta \vert F_\phi\vert$ increases approximately linearly with $\kappa,$ as shown in Fig.~\ref{fig:f_phi}(d). 
The values for the slope of a linear fit are shown as an inset in the figure. 
The fact that the chiral force decreases with $|\ell|$ is confirmed by the experimental data as discussed below. 

When taking Eq.~\eqref{eq:T} into account, $\Delta (\vert F_\phi\vert/P)$ leads to a difference between the orbital periods when the chiral beads are trapped with positive and negative laser beams. Such difference is the main experimental observable employed to characterize the chiroptical response of individual chiral particles. 
It is important to emphasize that the achiral contribution to the total optical force cancels out in $\Delta(\vert F_\phi\vert /P)$. 
Consequently, this observable isolates the chiral contribution to the force from any achiral background. 
Direct enantiocharacterization of the chiroptical response of individual chiral particles therefore becomes possible without requiring arbitrarily large values of $\kappa$.
This contrasts with most existing enantioselective techniques based on optical forces~\cite{martinez2024longitudinal}.

While the theoretical results shown in Fig.~\ref{fig:f_phi} are derived for the ideal case of laser beams with zero astigmatism, optical systems in a real experiment are inherently subject to aberrations that break the rotational symmetry about the optical $z$-axis. 
For instance, misalignment of the optical setup can introduce astigmatism. 
In the presence of astigmatism, the ring of maximal intensity is elongated along a specific direction.
Hence, the trapped bead does not strictly follow a circular orbit but rather an elliptical one.
While the SLM is employed to compensate for optical aberrations, a residual astigmatism remains and should be accounted for in the theoretical model to obtain quantitative agreement with the experimental data.    
To account for these optical aberrations, we employ the Mie-Debye Spherical Aberration+Astigmatism (MDSA+) theory \cite{Dutra2014} to compute the optical force on a trapped bead.   
Our model builds on previous extensions of the Mie-Debye theory for chiral particles trapped by Gaussian~\cite{Ali2020,ali2020probing} or vortex beams~\cite{Diniz2025}.

\begin{figure}
\centering
\tabskip=0pt
\valign{#\cr
    \hbox{%
    \begin{minipage}[b]{.43\linewidth}
\centering
\begin{tikzpicture}
\node[inner sep=0] (img)
  {\includegraphics[width=\linewidth]{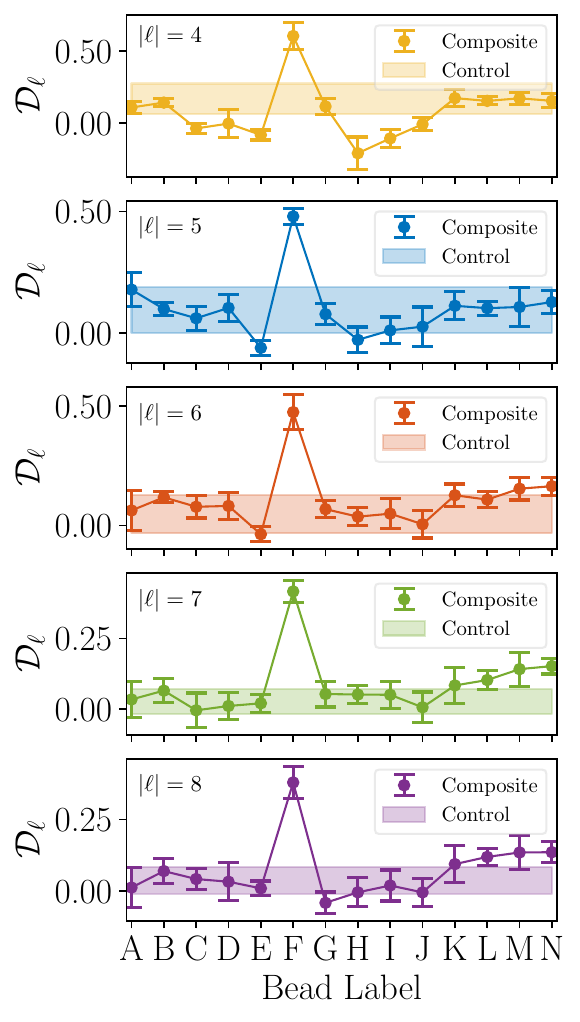}};
\node[
  anchor=north west,
  font=\scriptsize
] at ([xshift=5pt,yshift=7pt] img.north west) {(a)};
\end{tikzpicture}
\end{minipage}
  }\cr
    \noalign{\hfill}
  \hbox{%
    \begin{minipage}{.54\linewidth}
\centering
\begin{tikzpicture}
\node[inner sep=0] (img)
  {\includegraphics[width=\linewidth]{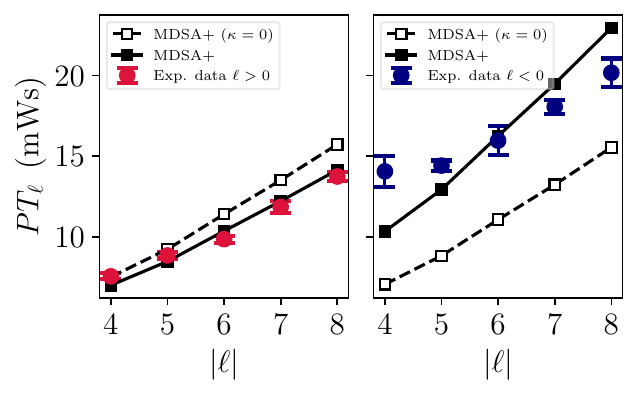}};
\node[
  anchor=north west,
  font=\scriptsize
] at ([xshift=0pt,yshift=6pt] img.north west) {(b)};
\end{tikzpicture}
\end{minipage}
  }\vfill
  \hbox{%
\begin{minipage}{.54\linewidth}
\centering
\begin{tikzpicture}
\node[inner sep=0] (img)
  {\includegraphics[width=\linewidth]{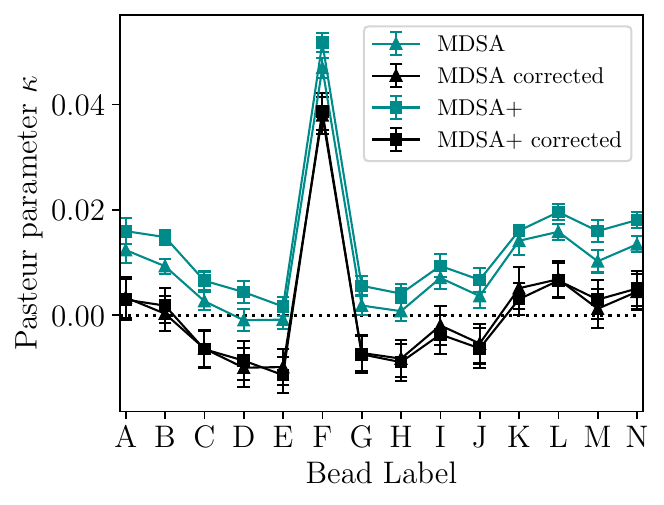}};
\node[
  anchor=north west,
  font=\scriptsize
] at ([xshift=0pt,yshift=2pt] img.north west) {(c)};
\end{tikzpicture}
\end{minipage}
  }
\cr
}

\caption{
(a) Topological dissymmetry factor $\mathcal{D}_\ell$ for  composite particles labeled from A to N. We consider topological charges $|\ell|=4,...,8$ from top to bottom. 
The shaded regions represent the average dissymmetry factor found for the control beads in each mode, defining a range where composite beads are indistinguishable from achiral control ones. 
(b) Orbital period (multiplied by the laser power) versus topological charge for bead F: experimental (circle with error bars) and 
MDSA+ fit with (filled square) or without (empty square) chirality. We use two fitting parameters for both models (see text for details).  
(c) Fitted Pasteur parameter $\kappa$ derived within the MDSA (triangle) and MDSA+ (square) theories. The black symbols show the corrected Pasteur parameters after subtracting the corresponding offsets in both theories.
}
\label{fig:fitting_results}
\end{figure}
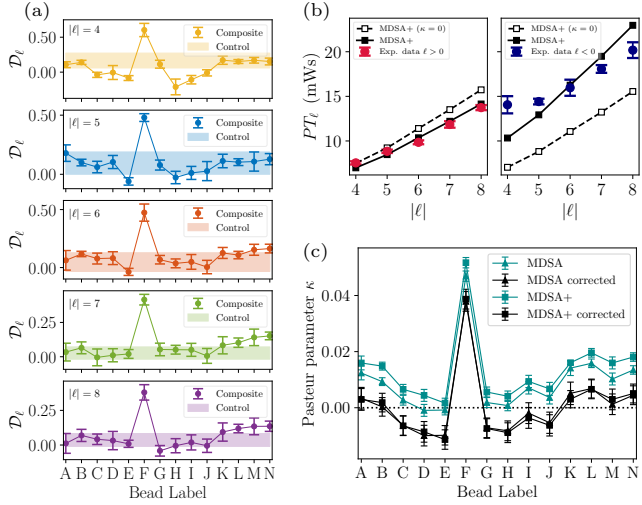

Experimentally, it is challenging to directly determine the astigmatism parameters quantitatively. 
However, these parameters can be obtained indirectly~\cite{Otsu-Hyodo2024}. 
Here this is achieved by using the measurements for the control beads.
By measuring the stable orbits $\rho_\text{eq}(\phi)$ of these control beads in a focused vortex beam, the astigmatism parameters can be extracted by fitting the experimental data to the MDSA+ theory (see SM~\cite{supplement} for details). 
We then model the experimental orbital periods of the composite beads taking the same astigmatism parameters found from the control beads, with the filling fraction of TiO$_2$ nanoparticles and the Pasteur parameter of the shell as the only fitting parameters. 

\begin{figure*}[htbp]
\centering

\begin{tikzpicture}
\node[inner sep=0] (img)
  {\includegraphics[width=1\linewidth, trim={0cm 0cm 2cm 0cm},
        clip]{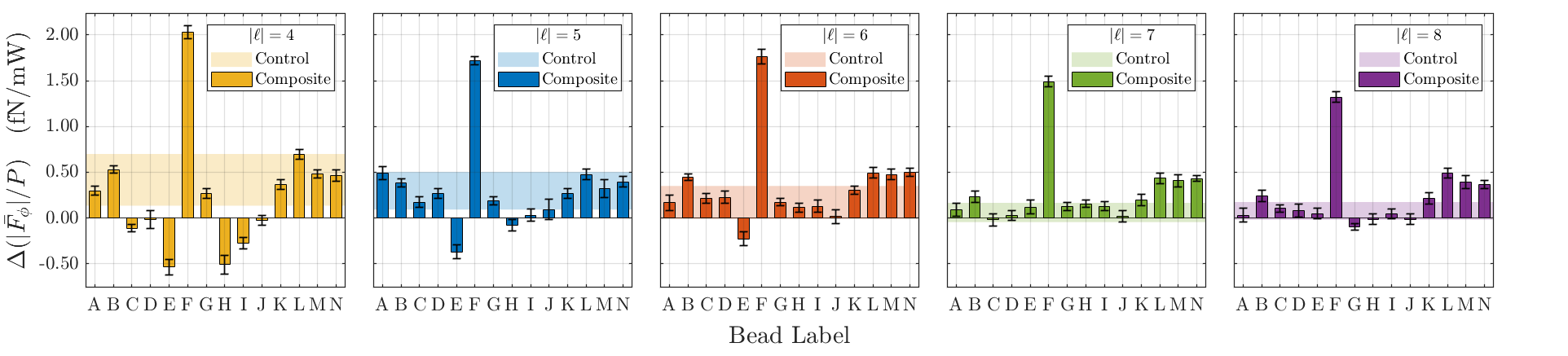}};
\node[anchor=north west, font=\scriptsize] 
  at ([xshift=30pt,yshift=-5pt] img.north west) {(a)};
  \node[anchor=north west, font=\scriptsize] 
  at ([xshift=127pt,yshift=-5pt] img.north west) {(b)};
    \node[anchor=north west, font=\scriptsize] 
  at ([xshift=224pt,yshift=-5pt] img.north west) {(c)};
      \node[anchor=north west, font=\scriptsize] 
  at ([xshift=322pt,yshift=-5pt] img.north west) {(d)};
        \node[anchor=north west, font=\scriptsize] 
  at ([xshift=419pt,yshift=-5pt] img.north west) {(e)};
\end{tikzpicture}

\caption{
Difference in azimuthal optical force normalized by the power $\Delta (|\bar{F}_{\phi}| / P)$ observed for each composite bead (solid bars).
$\Delta (|\bar{F}_{\phi}|/P)$ represents twice the chiral optical force as determined by comparing the orbital periods for positive and negative beams. 
From left to right, different panels show the results for topological charges from  $|\ell| = 4$ to $|\ell| = 8$, with the corresponding error bars for their mean standard deviation.
As in Fig.~\ref{fig:fitting_results}, the shaded region represents the average force difference for the control beads in each mode, and can be interpreted as an offset value for the chiral force.}
\label{fig:chiral_optical_force}
\end{figure*}

{\it Results and discussion---} We define the topological dissymmetry factor as
\begin{equation}
    \mathcal{D}_{\ell} = \frac{\tilde{T}_{-|\ell|} - \tilde{T}_{+|\ell|}}{(\tilde{T}_{-|\ell|} + \tilde{T}_{+|\ell|})/2},
    \label{eq:dl}
\end{equation}
which characterizes the chiroptical response of the composite bead trapped by the vortex laser beam, inspired by an analog description of helical dichroism for vortex beams~\cite{Ni2021, Ni2021-2}. 
$\mathcal{D}_{\ell}$ isolates the mechanical effect of a chiral force component, with $\tilde{T}_{\ell} = P T_{\ell}$ representing the orbital period $T_\ell$ scaled by the corresponding beam power. 
Importantly, in Eq.~\eqref{eq:dl} the contribution of the achiral optical force to the orbital period cancels out, in principle rendering $\mathcal{D}_{\ell}$ directly proportional to the particle's Pasteur parameter $\kappa$, and thereby isolating its chiroptical response.
Indeed, the suppression of the achiral background force shown in Fig.~\ref{fig:f_phi} implies that $\mathcal{D}_{\ell}$ is primarily driven by the chiroptical response of the beads, apart from a residual effect arising from the difference in optical aberrations between the positive and negative beams.
 
The experimental results for each topological charge $|\ell|$ are depicted in Fig.~\ref{fig:fitting_results}(a), where 
the shaded areas represent the variation of the dissymmetry factor found for the control beads. 
These areas define ranges of $\mathcal{D}_{\ell}$ that correspond to achiral beads, hence defining both an offset and the sensitivity for the determination of nonzero $\kappa$ values. 

The experimental data for $\tilde{T}_{\ell}$ (circles with error bars) together with the best MDSA+ fits are shown in Fig.~\ref{fig:fitting_results}(b) for one of the composite particles.  
We take the Pasteur parameter and the filling fraction of the shell as adjustable parameters (filled squares) and obtain excellent agreement with the experimental results for positive topological charges.
The values of $\kappa$ found for each individual bead are shown in Fig.~\ref{fig:fitting_results}(c). In addition to the MDSA+ results, we also show the values found without taking astigmatism into account (MDSA). 
Bead F is found to exhibit the largest value of $\kappa \approx 0.04$.
More details on the fitting parameters are presented in SM~\cite{supplement}.

To ensure that the computed Pasteur parameters do not result from artifacts of the fitting procedure or other effects such as a larger filling fraction or shell thickness, we set the chirality parameter to zero and use the filling fraction and shell thickness as fitting parameters for the experimental data corresponding to bead F; the resulting fit is also shown in Fig.~\ref{fig:fitting_results}(b) (empty squares). 
For $\kappa = 0$, the obtained periods for positive and negative topological charges are nearly identical, apart from small variations arising from the different astigmatism parameters. 
Consequently, the achiral model is unable to reproduce the experimentally observed difference between the periods associated with positive and negative topological charges within the experimental error bars.

However, the values for the Pasteur parameters obtained from the fit also capture in part a small asymmetry between the positive and negative vortex beams employed in the experiment. 
Such asymmetry is apparent in the range of the dissymmetry factors found for the control beads shown in Fig.~\ref{fig:fitting_results}(a), where negative beams lead to slower orbital motions (i.e. larger periods) for all values of $|\ell|$ as illustrated by Fig.~\ref{fig:fitting_results}(b). 
In order to quantify such asymmetry, we fit the data for the control beads using the chiral core-shell model.  
The resulting Pasteur parameters are small, but exhibit a systematic shift toward positive values (see SM~\cite{supplement}). 
This offset arises from an asymmetry between the positive and negative beams, possibly due to residual aberrations that are not yet considered in our theoretical model and affect the negative beams more severely. 
The resulting bias is accounted for by subtracting the offset value, given by the average Pasteur parameter for the control beads, from the results for the composite particles. 
After averaging over all control beads, we find the offset $\bar{\kappa}_\text{MDSA+} = 0.013 \pm 0.003$ and $\bar{\kappa}_\text{MDSA}= 0.009 \pm 0.003$, with the latter obtained without taking astigmatism into account. 
These offsets were subsequently used to correct the Pasteur parameters obtained for the composite particles as shown in Fig.~\ref{fig:fitting_results}(c) (black symbols). 
After subtracting the offset value, the corrected Pasteur parameters are consistent with the qualitative analysis of the dissymmetry factor shown in Fig.~\ref{fig:fitting_results}(a). 
In addition, we find similar numbers of right-handed and left-handed particles, as expected for a chiroptical response resulting from disorder. Some particles can be considered to be achiral as their 
fitted Pasteur parameters cannot be distinguished from zero within error bars. The errors of the fit typically yield values of $\delta\kappa\sim 10^{-3}$. This value therefore represents the 
sensitivity of our implementation, which is based on chiral optical forces exerted by vortex beams. 

As a final note on the characterization of $\kappa$, Fig.~\ref{fig:fitting_results}(c) demonstrates that the final corrected values obtained with and without accounting for astigmatism (MDSA+ and MDSA, respectively) are virtually indistinguishable. 
While modeling astigmatism is essential to achieve quantitative agreement with the experimental data for the period, the resulting values of $\kappa$ remain independent of it. 
This insensitivity to astigmatism parameters underscores the robustness of our characterization method.

Our method also allows us to determine the chiral optical force by taking the difference between the optical forces (per unit of power) $|\bar{F}_\phi|/P = 2\pi \beta \bar{\rho}_\text{eq}/(PT)$ found for positive and negative vortex beams (with bars denoting averages along the slightly elliptical orbits).
The results for $\Delta (|\bar{F}_{\phi}| / P) = |\bar{F}_{\phi,+|\ell|}|/P_+ - |\bar{F}_{\phi,-|\ell|}|/P_-$ for each bead are shown in Fig.~\ref{fig:chiral_optical_force}. 
As in Fig.~\ref{fig:fitting_results}(a), the shaded area represents an offset as determined from the variation found for the (achiral) control beads. Chiral forces per unit power in the range of fN/mW were found for bead F for all values of topological charge. 
Figure~\ref{fig:chiral_optical_force} demonstrates that the chiral force decreases monotonically with $|\ell|$, confirming the theoretical predictions shown in Fig.~\ref{fig:f_phi}(d).

\textit{Conclusions---}In summary, we have shown that disorder-induced chirality in TiO$_2$-coated silica microspheres leads to enhanced enantioselective optical forces and altered orbital dynamics in Laguerre–Gaussian optical tweezers. 
The proposed trapping scheme suppresses achiral contributions to the net force, enabling the direct observation of chiral optical forces acting on individual particles. 
A Mie–Debye model including optical aberrations quantitatively reproduces the measured dynamics and provides access to the effective Pasteur parameter of isolated microspheres. 
The large values obtained, reaching $\kappa\sim10^{-2}$, find their origin at the geometrical chirality associated with the random structure, establishing disordered all-dielectric composites as a versatile material platform for engineering chiral light–matter interactions and exploiting optical forces for chiral discrimination of single particles at the nanoscale.

\nocite{Otsu-Hyodo2024, Viana2006, giles2021,stober1968,
  Goldman1967, Schaffer2007,
  Viswanath1989,
  Pastoriza-Santos2004,Bohren1975,Garnett1904, Garnett1906,Markel2016,
  pinheiro2017spontaneous, Ali2020,
  lindell1994electromagnetic,
  Dutra2014,Fonseca2024, Diniz2024, 
  Daimon2007}

\bigskip
\textit{Acknowledgments---}We thank Arthur Fonseca, Gert-Ludwig Ingold, Jorge Olmos-Trigo, Luis Pires, Marcos Farina, and Rafael Dutra for fruitful discussions and assistance with various tasks related to the results reported herein. 
TS acknowledges funding from the Region Ile-de-France in the framework of the DIM QuanTiP. 
This work was partially supported by the Brazilian agencies Conselho Nacional de Desenvolvimento Científico e Tecnológico (CNPq) and Coordenação de Aperfeiçoamento de Pessoal de Nível Superior (CAPES), as well as the Research Foundation of the State of Rio de Janeiro (FAPERJ). 
This work was also supported by the CAPES-COFECUB collaboration project n° Ph1030/24 and by the Interdisciplinary Thematic Institute QMat, as part of the ITI 2021-2028 Program of the University of Strasbourg, CNRS and Inserm, IdEx Unistra (ANR 10 IDEX 0002), SFRI STRAT’US Project (ANR 20 SFRI 0012), and ANR-17-EURE-0024 under the framework of the French Investments for the Future Program. 
Support of USIAS (ANR-10-IDEX- 0002-02) and state funding from the ANR under the France 2030 program with reference ANR-23-EXLU-0004, PEPR LUMA TORNADO are also acknowledged.

\bibliography{references}

\end{document}

\title{Supplementary Material: Enantioselective optical trapping and characterization of all dielectric disorder-enabled chiral particles}

\author{Guilherme T. Moura}
 \email{guilhermeluis@pos.if.ufrj.br}
\affiliation{Instituto de F\'{\i}sica, Universidade Federal do Rio de Janeiro \\ Caixa Postal 68528,   Rio de Janeiro,  Rio de Janeiro, 21941-585, Brazil}
\affiliation{CENABIO - Centro Nacional de Biologia Estrutural e Bioimagem \\
Universidade Federal do Rio de Janeiro, Rio de Janeiro, Rio de Janeiro 21941-902, Brazil}
\author{Tanja Schoger}
\email{tanja.schoger@lkb.upmc.fr }
 \affiliation{Laboratoire Kastler Brossel, Sorbonne Université, CNRS, ENS-PSL, Collège de France, 75005 Paris, France}
 \author{Kainã G. Diniz}
\affiliation{Instituto de F\'{\i}sica, Universidade Federal do Rio de Janeiro \\ Caixa Postal 68528,   Rio de Janeiro,  Rio de Janeiro, 21941-585, Brazil}
\affiliation{CENABIO - Centro Nacional de Biologia Estrutural e Bioimagem \\
Universidade Federal do Rio de Janeiro, Rio de Janeiro, Rio de Janeiro 21941-902, Brazil}
\author{Marcel A. B. Morte}
\affiliation{Instituto de F\'{\i}sica, Universidade Federal do Rio de Janeiro \\ Caixa Postal 68528,   Rio de Janeiro,  Rio de Janeiro, 21941-585, Brazil}
\affiliation{CENABIO - Centro Nacional de Biologia Estrutural e Bioimagem \\
Universidade Federal do Rio de Janeiro, Rio de Janeiro, Rio de Janeiro 21941-902, Brazil}
\author{Diney S. Ether Jr}
\affiliation{Instituto de F\'{\i}sica, Universidade Federal do Rio de Janeiro \\ Caixa Postal 68528,   Rio de Janeiro,  Rio de Janeiro, 21941-585, Brazil}
\affiliation{CENABIO - Centro Nacional de Biologia Estrutural e Bioimagem \\
Universidade Federal do Rio de Janeiro, Rio de Janeiro, Rio de Janeiro 21941-902, Brazil}
\author{Leonardo de S. Menezes}
\affiliation{Chair in Hybrid Nanosystems, Nanoinstitute Munich, Faculty of Physics, Ludwig-Maximilians-University, 80539 Munich, Germany
}
\affiliation{
Departamento de Física, Universidade Federal de Pernambuco, 50670-901 Recife-PE, Brazil}
\author{Yicui Kang}
\affiliation{Chair in Hybrid Nanosystems, Nanoinstitute Munich, Faculty of Physics, Ludwig-Maximilians-University, 80539 Munich, Germany
}
\author{Emiliano Cort\'es}
\affiliation{Chair in Hybrid Nanosystems, Nanoinstitute Munich, Faculty of Physics, Ludwig-Maximilians-University, 80539 Munich, Germany
}
\author{Cyriaque Genet}
 \affiliation{Université de Strasbourg, CNRS, Institut de Science et d’Ingénierie Supramoléculaires, UMR 7006, F-67000 Strasbourg, France}
\author{Nathan B. Viana}
\email{nathan@if.ufrj.br}
\affiliation{Instituto de F\'{\i}sica, Universidade Federal do Rio de Janeiro \\ Caixa Postal 68528,   Rio de Janeiro,  Rio de Janeiro, 21941-585, Brazil}
\affiliation{CENABIO - Centro Nacional de Biologia Estrutural e Bioimagem \\
Universidade Federal do Rio de Janeiro, Rio de Janeiro, Rio de Janeiro 21941-902, Brazil}
\author{Felipe A. Pinheiro}
 \email{fpinheiro@if.ufrj.br }
\affiliation{Instituto de F\'{\i}sica, Universidade Federal do Rio de Janeiro \\ Caixa Postal 68528,   Rio de Janeiro,  Rio de Janeiro, 21941-585, Brazil}
\author{Paulo A. Maia Neto}
 \email{pamn@if.ufrj.br}
\affiliation{Instituto de F\'{\i}sica, Universidade Federal do Rio de Janeiro \\ Caixa Postal 68528,   Rio de Janeiro,  Rio de Janeiro, 21941-585, Brazil}
\affiliation{CENABIO - Centro Nacional de Biologia Estrutural e Bioimagem \\
Universidade Federal do Rio de Janeiro, Rio de Janeiro, Rio de Janeiro 21941-902, Brazil}%

\date{\today}% 

\maketitle

\tableofcontents

\section{Trapping and Acquisition Procedure}

We use tightly focused vortex beams to trap individual micro-sized beads. 
Specifically, we employ Laguerre-Gaussian modes $\mathrm{LG}_p^{|\ell|}$ with fixed radial number $p=0$ and topological charge $\ell$. 
By omitting the time propagation phase term, the electric field vector at the entrance of the objective can be written as
\begin{equation}
    \textbf{E}_\text{inc}(\mathbf{r}) = E_0 \left(\frac{\sqrt{2} \rho}{w_0}\right)^{|\ell|} e^{-\rho^2 /w_0^2} e^{ikz} e^{i\ell \phi} \left(\hat{x} + i \sigma \hat{y}\right)/\sqrt{2}
    \label{eq:lg_beam}
\end{equation}
where $\rho=\sqrt{x^2 +y^2}$, $w_0$ is the waist of the beam and $\sigma =1(-1) $ defines the left (right) circular polarized beams. 
The radius of maximum intensity for a given mode can be calculated by 
\begin{equation}
r_\ell = w_0 \sqrt{|\ell|/2}
\label{eq:rell}
\end{equation}
and the incident optical power is related to the field amplitude by $P_\text{in} = \frac{\pi}{4} \sqrt{\frac{\epsilon_{0}}{\mu_{0}}} w_{0}^{2} |E_{0}|^{2} |\ell|!$. 

By using a tightly focused Gaussian mode ($\ell = 0$), we trap a microsphere in the center of our camera's field of view.
Next, we lower the objective until the submicrometric particle touches the lower coverslip, a condition verified by an abrupt change in its Brownian agitation pattern. 
This procedure serves two purposes: registering the vertical position as $z = 0$ and ensuring that only one microsphere is in the trap. 
Subsequently, we displace the objective towards the sample by $(1.0 \pm 0.5)\,\microns$, establishing a standard trapping height to mitigate and control the influence of the spherical aberration introduced by refraction at the glass-water interface. 
The phase mask corresponding to the desired LG mode is loaded onto the SLM, and added to the phase compensation developed as a superposition of Zernike polynomials. 
The procedure to find the compensation mask is detailed in Ref.~\cite{Otsu-Hyodo2024}. 
For each topological charge $\ell$, the beam waist $w_0$ is adjusted according to Table~\ref{tab:param} to keep the objective's filling factor approximately constant, with $r_\ell/R_{\text{obj}} \approx 0.8$ ($R_{\text{obj}}$ is the radius of the objective entrance port). 

\begin{figure}
    \centering
    \includegraphics[width=0.2\linewidth]{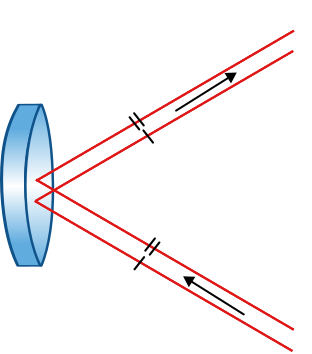}
    \caption{Schematic representation of parity transformation by a mirror. The two optical rays represent the ``edges'' of a beam. The incident ray with a dash, on the left of the beam, is found on the right after reflection. The reflection by the mirror effectively inverts the direction of rotation of the azimuthal phase ($e^{i\ell\phi} \rightarrow e^{-i\ell\phi}$).} 
    \label{fig:reflection}
\end{figure}

The generation of modes with negative values of $\ell$ is achieved by inserting a retractable mirror into the optical path of the positive ones, which effectively inverts the direction of the orbital angular momentum. 
Physically, the reflection performs a parity inversion on one of the transverse axes, as shown in Fig.~\ref{fig:reflection}. 
Under the transformation $x \to -x$ and $y \to y$, the azimuthal coordinate transforms as $\phi = \arctan(y / x) \to \arctan(-y / x) = -\phi$. 
From Eq.~\eqref{eq:lg_beam} we see that the azimuthal phase term $e^{i\ell\phi}$ is taken into $e^{-i\ell\phi}$, effectively inverting the wavefront helicity. 
In other words, by putting a mirror along the beam's optical path we are able to change the beams topological charge from $\ell$ to $-\ell$ in a purely geometric way. It should be noted that the beam polarization at this stage is linear, so the mirror has no effect on it.
After the topological charge is determined, we set the beam polarization state to circular by using a quarter-wave plate. 
We choose the polarization such that the optical spin and orbital angular momentum have the same sign, \textit{i.e.}, left circular polarization for modes with $\ell > 0$ and right circular polarization for $\ell < 0$. 
This ensures that the two experimental configurations are opposite to each other, so that a particle with a certain chirality responds more effectively to one configuration than the other, while an achiral particle will respond equally to both configurations.

\begin{figure}
    \centering
    \includegraphics[width=0.5\linewidth]{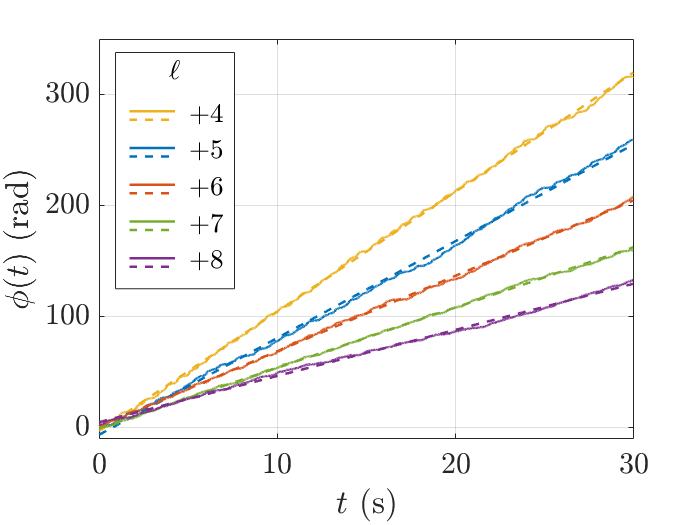}
    \caption{Time evolution of the unwrapped azimuthal coordinate $\phi(t)$ for composite bead A. 
    Solid lines: Experimental curve obtained over a 30-second segment. 
    Dashed lines: Linear fit. Both curves are plotted for modes $\ell = +4, \ldots, +8$ with left circular polarization. 
    The angular velocity $\omega$ obtained in each case is: $\omega_{\ell = +4} = (10.769 \pm 0.004)\,\text{rad/s}$, $\omega_{\ell = +5} = (8.720 \pm 0.007)\,\text{rad/s}$, $\omega_{\ell = +6} = (6.791 \pm 0.003)\,\text{rad/s}$, $\omega_{\ell = +7} = (5.431 \pm 0.002)\,\text{rad/s}$, $\omega_{\ell = +8} = (4.156 \pm 0.004)\,\text{rad/s}$.} 
    \label{fig:phiplot}
\end{figure}

For each mode, the orbital motion of a given particle was recorded for 3 minutes at 100 frames per second. 
On a single measurement run, we divide each 3-minute video into six 30-second video segments, accompanied by their corresponding power measurements. 
To extract the particle's angular velocity, a linear fit is performed on the angular displacement curve for each 30-second segment, given by
\begin{equation}
    \phi_\text{fit}(t) = \omega t + \phi_0\,.
\end{equation}
The slope of such a fit corresponds directly to the orbital angular velocity $\omega$. 
Figure~\ref{fig:phiplot} illustrates the method, showing experimental curves for the angular displacement $\phi_\text{exp}$ (solid lines) and the corresponding linear fits $\phi_\text{fit}$ (dashed lines) for modes $\ell = +4 ~\text{to}+8$. 
Finally, the measured orbital period is determined as
\begin{equation}
    T_k = \frac{2\pi}{\langle |\omega|\rangle_k}\,,
\end{equation}
where $k$ maps each period measurement to one of the six 30-second sections. 
By propagating the error of the slope ($\sigma_\omega$) obtained from the fit, the standard deviation of the period is given by $\sigma_T = (2\pi/\langle|\omega|\rangle^2) \sigma_\omega$. 
Thus, the period measurement of a particle in a given spin-orbit mode is
\begin{equation}
    T = \bar{T} \pm \frac{\sigma_T}{\sqrt{N}}\,,
\end{equation}
where $N = 6$ is the number of realizations and $\bar{T}$ represents the arithmetic mean.

The videos are first acquired for positive values of the topological charge ($\ell = +4, +5, +6, +7, +8$) with left-circular polarization. 
After lowering the retractable mirror and adjusting the wave plates properly, we acquire the videos for negative modes ($\ell = -4, -5, -6, -7, -8$) with right-circular polarization. 
The power in the sample region, crucial for comparing theory and experiment, is given by $P=T_\text{obj}P_\text{in}$, where $T_\text{obj}$ is the objective transmittance for each mode. We measure $T_\text{obj}$ following the method described in Ref.~\cite{Viana2006}. The values of $T_\text{obj}$ and of the power $P_\text{in}$ at the entrance of the objective are presented in Table~\ref{tab:param}. 

For a fixed beam waist $w_0$, the radius of maximum intensity before focusing $r_\ell$ increases with $\vert{}\ell\vert{}$. Therefore, to maintain an approximately uniform filling factor at the objective aperture across different modes, $w_0$ must be reduced as $\vert{}\ell\vert{}$ increases. We measure the experimental width $w_0$ of the beam for each mode with the help of a beam profiler and then calculate $r_{\ell}$ from Eq.~\eqref{eq:rell}. The resulting values are shown in Table~\ref{tab:param}. In summary, the beam parameters compiled in Table~\ref{tab:param} across various modes are crucial to accurately modeling the optical forces and the resulting orbital motion of the trapped particles.

\begin{table}[h] 
 \renewcommand{\arraystretch}{1.1}
    \setlength{\tabcolsep}{5pt}
\begin{tabular}{c|c|c|c|c|c}
$|\ell|$ & $w_0$ (mm) & $r_\ell$ (mm) & $T_\text{obj}$ (\%) & $P_{\text{in},\ell > 0}$ (mW) & $P_{\text{in},\ell < 0}$ (mW) \\
\hline
4 & 1.46 $\pm$ 0.03 & 2.07 $\pm$ 0.05 & 32.71 & 70.56 $\pm$ 0.01 & 75.00 $\pm$ 0.01 \\
5 & 1.36 $\pm$ 0.01 & 2.15 $\pm$ 0.01 & 32.84 & 65.00 $\pm$ 0.01 & 69.00 $\pm$ 0.01 \\
6 & 1.25 $\pm$ 0.02 & 2.16 $\pm$ 0.03 & 33.20 & 60.56 $\pm$ 0.01 & 62.00 $\pm$ 0.01 \\
7 & 1.19 $\pm$ 0.02 & 2.23 $\pm$ 0.03 & 33.15 & 57.22 $\pm$ 0.01 & 59.00 $\pm$ 0.01 \\
8 & 1.14 $\pm$ 0.01 & 2.28 $\pm$ 0.02 & 33.16 & 53.89 $\pm$ 0.01 & 56.00 $\pm$ 0.01 
\end{tabular}
\caption{Experimental beam parameters.}
\label{tab:param}
\end{table}

\section{Sample Preparation and Characterization}

Two classes of beads were used: silica (SiO$_2$) spheres (control beads, achiral) and silica spheres of the same size as the control, but coated with titanium dioxide (TiO$_2$, anatase phase) nanoparticles of approximately 10\,nm in diameter, which are the origin of their geometric chirality. 
The synthesis of both classes of particles were adapted from previously reported procedures with modifications~\cite{giles2021,stober1968}, according to the following procedures. 

\subsection{Materials}

Tetraethyl orthosilicate (TEOS, 98\%), ammonium hydroxide solution (NH$_4$OH), poly(allylamine hydrochloride) (PAH), sodium chloride, and sodium citrate were purchased from Sigma-Aldrich and used as received. Anatase TiO$_2$ nanoparticles ($<25$ nm) were also obtained from Sigma-Aldrich.

\subsection{Preparation of SiO$_2$ Particles}

Monodisperse SiO$_2$ spheres with a nominal diameter of 600\,nm were synthesized using a modified Stöber method~\cite{stober1968}. 
Briefly, 18.12\,mL of absolute ethanol, 3.21\,mL of deionized (DI) water, and 1.96\,mL of NH$_4$OH solution were mixed and stirred at 800\,rpm at room temperature. 
Subsequently, 1.7\,mL of TEOS was added to the reaction mixture. 
The suspension became turbid within a few minutes, indicating the formation of SiO$_2$ particles. 
The resulting SiO$_2$ spheres were purified by three centrifugation--redispersion cycles in ethanol.

\subsection{Deposition of TiO$_2$ Nanoparticles onto SiO$_2$ Particles}

TiO$_2$ nanoparticles (100\,mg) were dispersed in 10\,mL of DI water. 
A 4\,mL aliquot of this suspension was added to 200\,mL of 0.1\,M sodium citrate solution and stirred overnight to obtain citrate-functionalized TiO$_2$ nanoparticles.

For surface functionalization of the SiO$_2$ particles, 200\,$\upmu$L of the SiO$_2$ suspension was added to 10\,mL of PAH solution prepared in 0.5\,M sodium chloride solution (pH 5.0). 
The mixture was stirred for 1\,h at 800\,rpm, after which the PAH-coated SiO$_2$ particles were collected by centrifugation, washed three times with DI water to remove excess PAH, and redispersed in DI water to a final volume of 15\,mL.

To obtain the chiral composite beads composed of SiO$_2$ decorated with TiO$_2$ nanoparticles with different TiO$_2$ surface coverages, 10, 25, or 50\,mL aliquots of the citrate-functionalized TiO$_2$ suspension were separately collected, washed three times with DI water by centrifugation, and each redispersed in 10\,mL of DI water. 
Subsequently, 5\,mL of the PAH-coated SiO$_2$ suspension was added to each TiO$_2$ suspension under continuous stirring at 600\,rpm for 1.5\,h to promote electrostatic adsorption of TiO$_2$ nanoparticles onto the SiO$_2$ surface. 
The resulting composite SiO$_2$ decorated with TiO$_2$ particles were washed once with DI water by centrifugation to remove unbound TiO$_2$ nanoparticles and stored in DI water until further use. 

\begin{figure}
    \centering
    \includegraphics[width=1\linewidth]{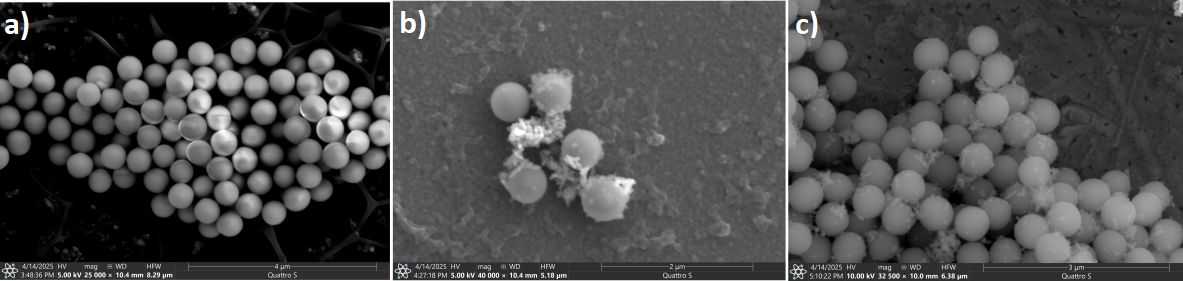}
    \caption{Scanning electron microscopy (SEM) images of the beads used in the experiments: (a) control beads, and (b), (c) composite beads.} 
    \label{fig:CENABIO_beads}
\end{figure}

Scanning electron microscopy (SEM) images of the particles are presented in Fig.~\ref{fig:CENABIO_beads}, where the presence of TiO$_2$ nanoparticles adhered to the surface of the SiO$_2$ spheres is clearly observed, while the control spheres present a smooth surface. 
Furthermore, through the SEM images, it was possible to measure the diameter of the silica spheres, for which we found the value of $(550 \pm 12) \,\mathrm{nm}$. 

\subsection{Determination of the average and maximum shell thicknesses}\label{sec:hbar_hmax}

\begin{figure}
    \centering
    \includegraphics[width=1\linewidth]{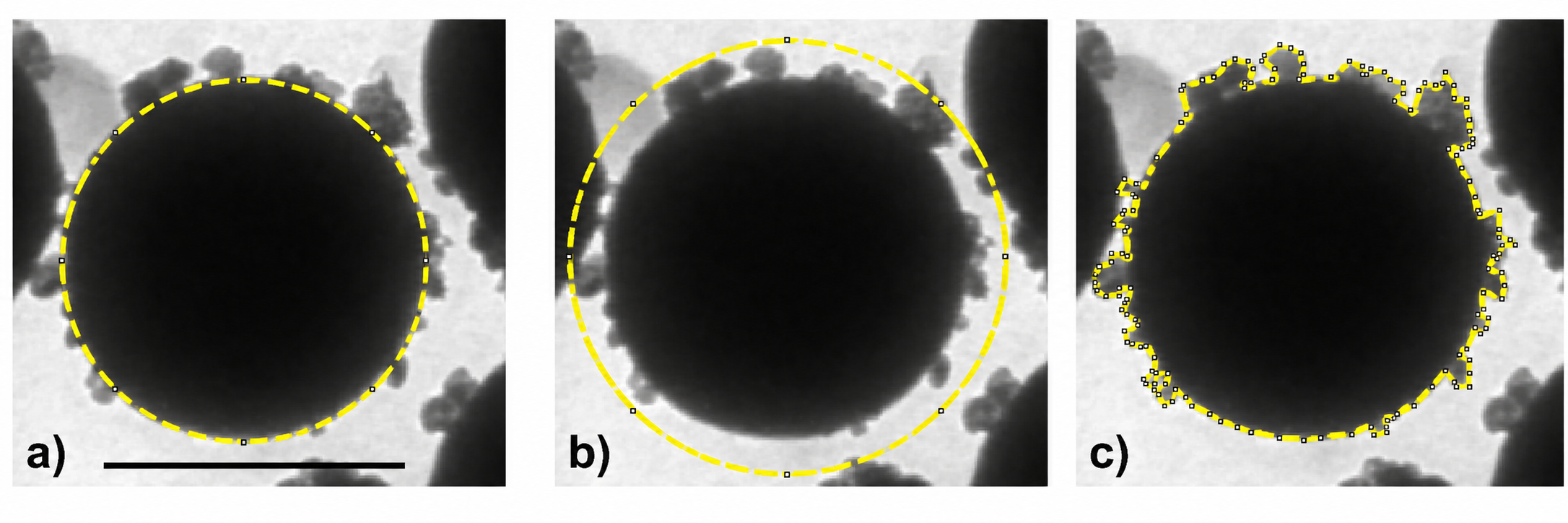}
    \caption{Transmission electron microscopy (TEM) images used to determine the composite parameters $\bar{h}$ and $h$. Yellow lines define the edges of the surfaces that determine a) $A_\text{core}$, b) $A_\text{max}$ and c) $A_\text{inc}$. Scale bar: $500\,\mathrm{nm}$.}
    \label{fig:hbar}
\end{figure}

We employed transmission electron microscopy (TEM) images of composite spheres to quantify the average ($\bar{h}$) and maximum ($h$) shell thicknesses. 
The quantification is based on three areas, obtained from a sphere image, as shown in Fig.~\ref{fig:hbar}.

$A_\text{core}$ is the area of the disk that defines the radius $R_\text{core}$ of the silica core. $A_\text{max}$ is the area of the disk of radius $a_\text{max}$ that encompasses all TiO$_2$ nanoparticles adhered to the surface of the silica microsphere. Both circles are defined with the same center. $A_\text{inc}$ is the area of the region containing just the nanoparticles together with the silica core (Fig.~\ref{fig:hbar} c).

Assuming that the local height of the composite with respect to the core surface $h(\theta,\phi)$ depends only on the azimuthal spherical coordinate $\phi$, the average shell thickness is given by:
\begin{equation} 
\bar{h}=\frac{\int_0^{2\pi}d\phi\int_0^{\pi}d\theta\, h(\phi)  \sin(\theta) }{4 \pi }.
\label{eq:hbar_eq1}
\end{equation}
Expressing $\bar{h}$ exclusively as a function of the measured areas, we find
\begin{equation} 
\bar{h}=\frac{\int_0^{2 \pi} h(\phi) R_\text{core} d\phi}{2 \pi R_\text{core}}= \frac{A_\text{inc}-A_\text{core}}{\sqrt{4 \pi A_\text{core}}}.
\label{eq:hbar_eq2}
\end{equation}

The shell thickness ($h$) is defined as the maximum value of $h(\phi)$. In terms of the measured areas, it reads
\begin{equation}
h=a_\text{max}-R_\text{core}=\frac{\sqrt{A_\text{max}}-\sqrt{A_\text{core}}}{\sqrt{\pi}}.
\label{eq:hbar_eq3}
\end{equation}
From the analysis of the images of eight different composite particles and using the image scale bar for calibration, we obtain $\bar{h}=(23\pm 13)\,\mathrm{nm}$ and $h=(63\pm 17)\,\mathrm{nm}$. The uncertainties correspond to the standard deviation of the mean.

\section{Drag coefficient and Faxén correction}

\begin{figure}[b]
    \centering
    \begin{minipage}{0.45\linewidth}
    \includegraphics[width=1\linewidth]{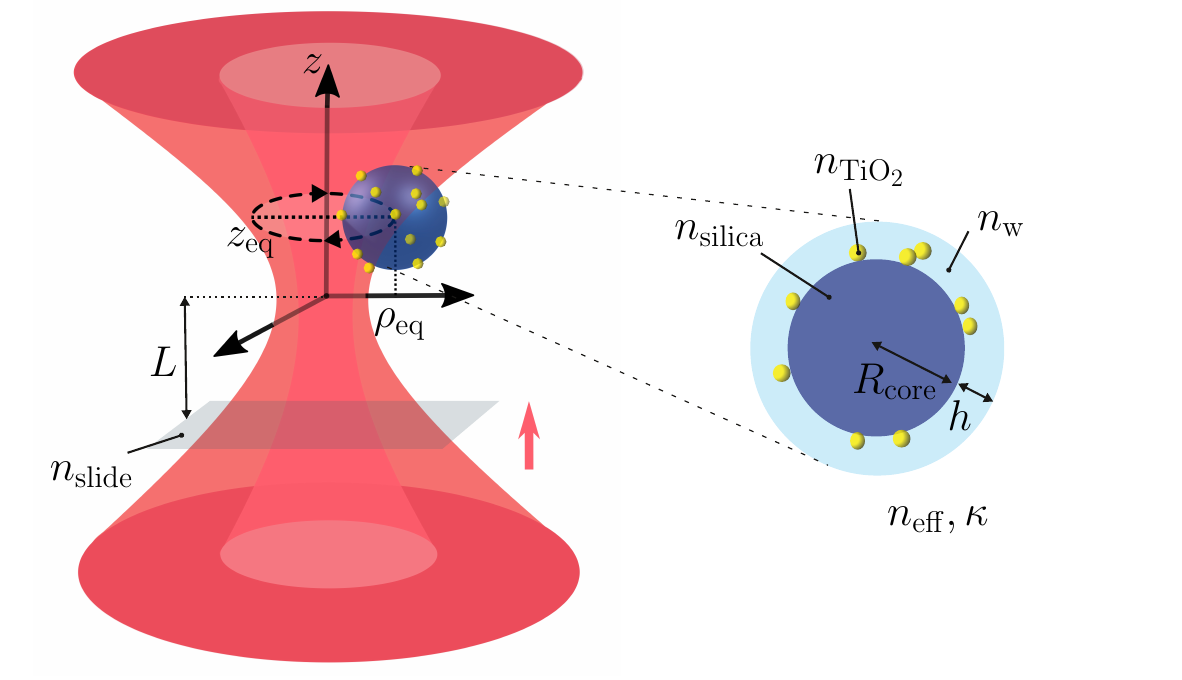}
    \end{minipage}
    \begin{minipage}{0.45\linewidth}
    \renewcommand{\arraystretch}{1.2}
    \setlength{\tabcolsep}{4pt}
        \begin{tabular}[t]{l}
        sphere parameters \\ \hline
            $R_\text{core}= 0.275 \,\microns$   \\
            $h = 0.063\,\microns $ \\
              $n_\text{silica} = 1.4146$ \\
              $n_\mathrm{TiO_2}=2.4789$ \\
                $n_\text{w} = 1.3242$ 
        \end{tabular}
        \hspace{2em}
        \begin{tabular}[t]{l}
        beam parameters \\ \hline
         $\lambda = 1.064\, \microns$ \\
         NA = 1.4 \\
         $R_\text{obj} = 2.8\,\microns$ \\
                $L = 0.87 \,\microns$ \\
                $n_\text{slide}= 1.518$ \\
        \end{tabular}
        \end{minipage}
            \caption{Schematic representation of the trapping geometry together with all relevant parameters.}
    \label{fig:parameters}
\end{figure}

The $\beta$ defines the viscous drag experienced by a spherical particle of radius $R$ moving in a fluid near a planar boundary. 
Including the Faxén correction \cite{Goldman1967, Schaffer2007}, the drag coefficient is given by
\begin{equation}
\beta =
\frac{6\pi \eta R}{
1
- \frac{9}{16}\frac{R}{d}
+ \frac{1}{8}\left(\frac{R}{d}\right)^3
- \frac{45}{256}\left(\frac{R}{d}\right)^4
- \frac{1}{16}\left(\frac{R}{d}\right)^5
}\,,
\label{eq:faxen_drag}
\end{equation}
where $\eta=1.018\,\mathrm{m Pa \cdot s}$ \cite{Viswanath1989} is the dynamic viscosity of water at room temperature and $d$ is the distance between the center of the particle and the glass slide at the bottom of the sample. 
For the composite particles, we take $R=R_{\rm core}+\bar{h},$ where $R_{\rm core}=(275\pm 6)\,$nm is the radius of the silica spherical particles and 
$\bar{h}=(23\pm 13)\,{\rm nm}$
is the average height of the shell as determined from the SEM images of the composites (see Sec.~\ref{sec:hbar_hmax}).
The distance $d= L+ z_\text{eq}$ is determined from $z_{\rm eq}$, the axial equilibrium with respect to the focus, and from the height $L$ of the focal plane with respect to the glass slide (see Fig.~\ref{fig:parameters}).

\section{Effective model description of the dielectric permittivity of the composite chiral particles}\label{sec:maxwell_garnett}

To describe the optical properties of the coated composite particles, we employ an effective-medium approach \cite{Pastoriza-Santos2004}. 
The composite particles are modeled using a core–shell geometry~\cite{Bohren1975}, where the chiral shell consists of the TiO$_2$ nanoparticles dispersed in water (see Fig.~\ref{fig:parameters}). 
The effective refractive index $n_\text{eff} = \sqrt{\epsilon_\text{eff}}$ of the shell is calculated using the Maxwell–Garnett model \cite{Garnett1904, Garnett1906,Markel2016}
\begin{equation}
\epsilon_{\text{eff}}= 
\epsilon_\text{w}
\frac{
\epsilon_{\text{TiO}_2} + 2\epsilon_\text{w} + 2f(\epsilon_{\text{TiO}_2}- \epsilon_\text{w})
}{
\epsilon_{\text{TiO}_2} + 2\epsilon_\text{w} - f(\epsilon_{\text{TiO}_2} - \epsilon_\text{w})
}\,,
\end{equation}
where $\epsilon_\text{w}=n_\text{w}^2$ denotes the permittivity of the host medium (water) and $\epsilon_{\text{TiO}_2}$ the permittivity of the nanoparticles, with $f \in [0,1]$ being their volume filling fraction. 
Owing to the random spatial arrangement of the nanoparticles that lacks mirror symmetry, the optical response of the shell can be chiral~\cite{pinheiro2017spontaneous, Ali2020}.
The constitutive equations for the chiral shell are given by \cite{lindell1994electromagnetic}
    \begin{align}
        \mathbf{D}(\mathbf{r},\omega) &= \epsilon_0 \epsilon_\text{eff}\mathbf{E}(\mathbf{r},\omega) + i\kappa \mathbf{H}(\mathbf{r},\omega)/c\,,
        \\
        \mathbf{B}(\mathbf{r},\omega) &= -i \kappa\mathbf{E}(\mathbf{r},\omega)/c + \mu_0 \mathbf{H}(\mathbf{r},\omega)
    \end{align}
where $\kappa$ is the Pasteur chirality parameter, $c$ is the speed of light in vacuum, and $\epsilon_0$ and $\mu_0$ denote the vacuum permittivity and permeability, respectively.

We fit the filling fraction $f$ and the Pasteur parameter $\kappa$ that characterizes the chiroptical response of the shell against the data for the orbital period $T$. As discussed in Sec.~\ref{sec:hbar_hmax}, we determine the shell thickness $h=(63\pm 17)\,{\rm nm}$ from the TEM images as the maximum height of the satellites with respect to the core surface after averaging over 8 composite particles (see Fig.~\ref{fig:hbar}).

\section{MDSA+ theory for chiral core-shell particles trapped with Laguerre-Gaussian beams}
 
Here, we present the series expansion of the optical force components for a tightly-focused Laguerre-Gaussian beam within the Mie Debye Spherical Aberration + Astigmatism (MDSA+) theory. 
The MDSA+ theory for various setups can already be found in the literature. 
More specifically, expressions for a tightly-focused Gaussian beam including optical aberrations can be found in Ref.~\cite{Dutra2014}, while extensions to Laguerre-Gaussian beams are presented in Refs.~\cite{Fonseca2024, Diniz2024}. 
The application of the MDSA theory (without astigmatism) for a Gaussian beam interacting with chiral core-shell particles can be found in Ref.~\cite{Ali2020}. 
Here, we combine all these results to apply the MDSA+ theory for a tightly-focused circular-polarized Laguerre-Gaussian beam interacting with a chiral core-shell particle. 
We define a dimensionless force $\mathbf{Q} = \mathbf{F}/(n_\text{w} P/c)$, with the laser beam power $P$, the speed of light $c$ and the refractive index in the sample region $n_\text{w}$. In our case, the particles are suspended in water, with $n_\text{w} = 1.3242$ \cite{Daimon2007}.
We include astigmatism and the spherical aberration introduced by refraction at the glass-water interface in the theoretical modeling. 

The optical force is separated into a scattering (s) and extinction (e) component ($ \mathbf{Q}(\mathbf{R}) = \mathbf{Q}_\text{s} + \mathbf{Q}_\text{e} $) and it is calculated at the position $\mathbf{R} =  \mathbf{R}(\rho, \phi, z)$ of the spherical particle with respect to the objective's focal point as depicted in Fig.~\ref{fig:parameters}. 
Each force component is defined for a certain circular polarization $\sigma= \pm 1$ and topological charge $\ell $ of the incoming trapping beam.  

The axial and transverse components of the scattering force in cylindrical coordinates are given by
\begin{equation}
\begin{aligned}
Q_{\text{s}z}^{(\sigma, \ell)} = -\frac{8\gamma^2}{A_\ell N_\text{s}} \mathrm{Re} 
\sum\limits_{j=1}^\infty \sum_{m=-j}^j 
&
\left\{
    \frac{\sqrt{j(j+2)(j-m+1)(j+m+1)}}{j+1} 
    \left[\alpha^\sigma_j (\alpha^\sigma_{j+1})^{*} + \beta^\sigma_j (\beta^\sigma_{j+1})^{*}\right]
    G_{j,m}^{(\sigma, \ell)} G_{j+1,m}^{(\sigma, \ell)*} \right. 
\\[7pt]
& \quad \left.
    + \sigma m \frac{2j+1}{j(j+1)} \alpha^\sigma_j (\beta^\sigma_j)^{*} \left|G_{j,m}^{(\sigma, \ell)}\right|^2 
    \right\}
\label{eq:Qsz}
\end{aligned}
\end{equation}
and 
\begin{equation}
\begin{aligned}
\left\{\begin{array}{c}
Q_{\text{s}\rho}^{(\sigma, \ell)} \\
Q_{\text{s}\phi}^{(\sigma, \ell)}
\end{array}\right\}
 = \frac{4\gamma^2}{A_\ell N_s} 
\left\{ \begin{array}{c}
 \mathrm{Im} \\
 - \mathrm{Re}
 \end{array}\right\}
 \sum\limits_{j=1}^\infty \sum_{m=-j}^j 
 & \left[\frac{\sqrt{j(j+2)(j+m+1)(j+m+2)}}{j+1} 
\left[\alpha^\sigma_j (\alpha^\sigma_{j+1})^{*} + \beta^\sigma_j (\beta^\sigma_{j+1})^{*}\right]
\right.
\\
&\hspace{5em} \times
\left( G_{j,m}^{(\sigma, \ell)} G_{j+1,m+1}^{(\sigma, \ell)*}
 \pm G_{j,-m}^{(\sigma, \ell)} G_{j+1,-m-1}^{(\sigma, \ell)*}\right)
\\[7pt]
&  \quad - 
\left. 2\sigma\frac{2j+1}{j(j+1)} \sqrt{(j-m)(j+m+1)} \mathrm{Re}\left[\alpha^\sigma_j (\beta^\sigma_j)^{*}\right]
 G_{j,m}^{(\sigma, \ell)} G_{j,m+1}^{(\sigma, \ell)*}
 \right]\,,
\end{aligned}
\label{eq:Qsrho}
\end{equation}
where the upper sign corresponds to the radial force component, while the lower sign is for the azimuthal part. 

The multipole coefficients are defined by 
\begin{equation}
\begin{aligned}
G_{j, m}^{(\sigma, \ell)}(\mathbf{R}) = (\sqrt{2}\gamma)^{|\ell|} 
\int_0^{\theta_\text{m}} \mathrm{d}\theta  
\sqrt{\cos(\theta)} \sin^{|\ell|+1}(\theta)
e^{-\gamma^2 \sin^2(\theta)}
d_{m, \sigma}^{j} (\theta_w) 
 f^{(\sigma, \ell)}_m (\mathbf{R})
e^{ik_\text{w}\cos(\theta) z + i\Psi_\text{g-w}} \,.
\end{aligned}
\label{eq:G}
\end{equation}
The phase factor $\Psi_\text{g-w} = k L\left(N_\text{s} \cos\theta_\text{w} 
		- \cos\theta/ N_\text{s}\right) $
accounts for the spherical aberration with $N_\text{s} = n_\text{w}/n_\text{slide}$ and $k= 2\pi n_\text{slide}/\lambda $ the wave number in the glass slide. 
$L$ is the distance between the glass slide and the objective's focal point, as shown in Fig.~\ref{fig:parameters}. 
Furthermore, the wave vector components in water are given by: $k_\text{w} = N_\text{s} k$  and $\sin\theta_w = \sin\theta/N_\text{s}$.
We integrate up to a maximal angle $\theta_\text{m}$ defined by $\sin\theta_\mathrm{m}= \mathrm{min}(N_\text{slide}, \mathrm{NA}/n_\text{slide})$.
$\gamma = f/w_0$ is the ratio of the focal length $f$ and the beam waist $w_0$. 
The values of the beam waist used for each topological charge are presented in Table~\ref{tab:param}.
The coefficient $f_m^{(\sigma, \ell)}$ accounts for astigmatism and is given by \cite{Diniz2024} 
\begin{equation}
\begin{aligned}
f^{(\sigma, \ell)}_m (\mathbf{R}) = &
\sum_{s=-\infty}^\infty (-i)^{s} J_s\left(2\pi A_\text{ast} \frac{\sin^2\theta}{\sin^2\theta_0}\right) 
J_{2s +m -\sigma-\ell}(k \rho \sin\theta) e^{2is(\phi_\text{ast} - \phi)}\,,
\end{aligned}
\end{equation}
where $J_n(z)$ are the Bessel functions of the first kind and $A_\text{ast}, \phi_\text{ast}$ are the astigmatism parameters. 
The factor $A_\ell$ accounts for the filling fraction of the beam and is given by
\begin{equation}
A_\ell = 8(2\gamma^2)^{|\ell|+1}
\int_0^{\sin\theta_\text{m}}  t^{2|\ell| +1} e^{-2\gamma^2 t^2}
\frac{\sqrt{(1-t^2)(N_\text{s}^2 - t^2)}}
	{\left(\sqrt{1-t^2} + \sqrt{N_\text{s}^2 - t^2}\right)^2}\,
	\mathrm{d}t \,.
\end{equation}

Finally, $\alpha^\sigma_j$ and $\beta^\sigma_j$ define the reflection coefficients of the spherical particle. For circular polarized beams, they are given by
\begin{align}
    \alpha_j^\sigma = a_j + i\sigma d_j,\quad
    \beta_j^\sigma = b_j - i\sigma c_j\,,
\end{align}
where $a_j, b_j, c_j$ and $d_j$ are the Mie scattering coefficients for the chiral core-shell sphere in the linear polarization case. They can be found in Ref.~\cite{Bohren1975}.
The Mie coefficients are defined by the core radius $R_\text{core}$, the shell thickness $h$, by the refractive indices of core ($n_\text{silica}$) and shell ($n_\text{eff}$), as well as the Pasteur parameter $\kappa$, which are all illustrated in Fig.~\ref{fig:parameters}. 
We employed the Maxwell-Garnett theory to describe the effective refractive index of the shell (see Sec.~\ref{sec:maxwell_garnett}). 

To complete the set of equations, the following relations hold for the extinction component of the optical force: 
\begin{equation}
Q_{\text{e}z}^{(\sigma, \ell)} = \frac{4\gamma^2}{A_\ell N_\text{s}}\mathrm{Re} \sum\limits_{j=1}^\infty \sum_{m=-j}^j (2j+1)(\alpha^\sigma_j + \beta^\sigma_j) 
G_{j,m}^{(\sigma, \ell)} \left(G^{(\sigma, \ell)'}_{j,m}\right)^{*}
\label{eq:Qez}
\end{equation}
with 
\begin{equation}
G^{(\sigma, \ell)'}_{j,m} = 
\frac{\sqrt{j(j+2)\left[(j+1)^2 -m^2 \right]}}{(2j+1)(j+1)}
G_{j+1, m}^{(\sigma, \ell)} 
+
\frac{\sqrt{(j^2 -m^2)(j^2-1)}}{j(2j+1)}  G_{j-1, m}^{(\sigma, \ell)} +
\sigma \frac{m}{j(j+1)} G_{l,m}^{(\sigma, \ell)}\,. 
\label{eq:G_prime}
\end{equation}

The transverse components of $\mathbf{Q}_\mathrm{e}$ are given by
\begin{equation}
\left\{\begin{array}{c}
Q_{\text{e}\rho}^{(\sigma, \ell)} \\
Q_{\text{e}\phi}^{(\sigma, \ell)}
\end{array}\right\} 
= \frac{2\gamma^2}{A_\ell N_\text{s}} 
\left\{\begin{array}{c}
\mathrm{Im} \\
- \mathrm{Re}
\end{array}\right\} 
 \sum\limits_{j=1}^\infty \sum_{m=-j}^j (2j+1)(\alpha^\sigma_j + \beta^\sigma_j) G_{j,m}^{(\sigma, \ell)}
  \left(G^{-, (\sigma, \ell)}_{j,m+1} \mp G^{+, (\sigma, \ell)}_{j,m-1}\right)^{*} \,,
\label{eq:Qerho}
\end{equation}
where the negative (positive) sign corresponds to the radial (azimuthal) component. 
The multipole coefficients $G^{\pm, (\sigma, \ell)}_{j,m}$ are given by
\begin{equation}
\begin{aligned}
G^{\pm, (\sigma, \ell)} _{j,m} &= 
\mp \frac{\sqrt{(j\pm m)(j \pm m+1)(j^2 -1)}}{j(2j+1)} G_{j-1,m}^{(\sigma, \ell)}  + 
\sigma \frac{\sqrt{(j\mp m)(j \pm m+1)}}{j(j+1)} G_{j,m}^{(\sigma, \ell)}  \\
& \qquad \pm
\frac{\sqrt{(j\mp m)(j\mp m+1)((j+1)^2 -1)}}{(j+1)(2j+1)} G_{j+1,m}^{(\sigma, \ell)} \,.
\end{aligned}
\label{eq:G_pm}
\end{equation}

\section{Determining the amount of astigmatism}\label{sec:fitting_Aast}

In this section, we describe how the astigmatism parameter $A_\text{ast}$ was inferred from the control particle dynamics. 
Astigmatism causes the focal spot to become elongated, which in turn leads the trapped particle to follow an elliptical rather than a circular orbit. 
By determining the bead's stable orbit and fitting our theoretical model to the measured trajectory, we can extract the magnitude of the astigmatism. 
For this fit, we are using the control measurement with the spherical dielectric beads, so without a shell. 

\begin{figure}[h]
    \centering
    \begin{subfigure}{.3\linewidth}
\centering
\begin{tikzpicture}
\node[inner sep=0] (img)
  {\includegraphics[height=4cm]{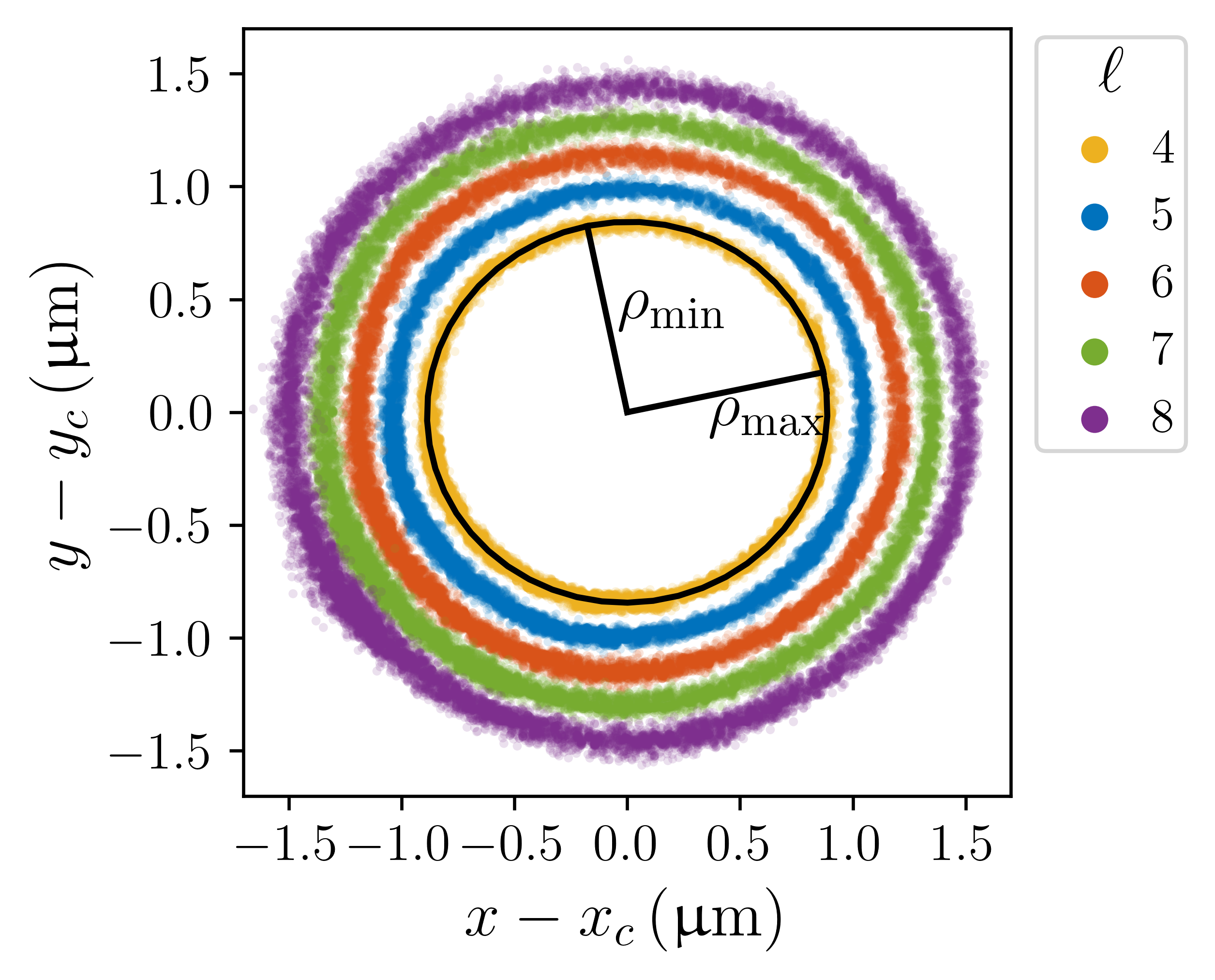}};
\node[
  anchor=north west,
  font=\scriptsize
] at ([xshift=5pt,yshift=11pt] img.north west) {(a)};
\end{tikzpicture}
\end{subfigure}
\hspace{0.5em}
    \begin{subfigure}{.3\linewidth}
\centering
\begin{tikzpicture}
\node[inner sep=0] (img)
  {\includegraphics[height=4cm]{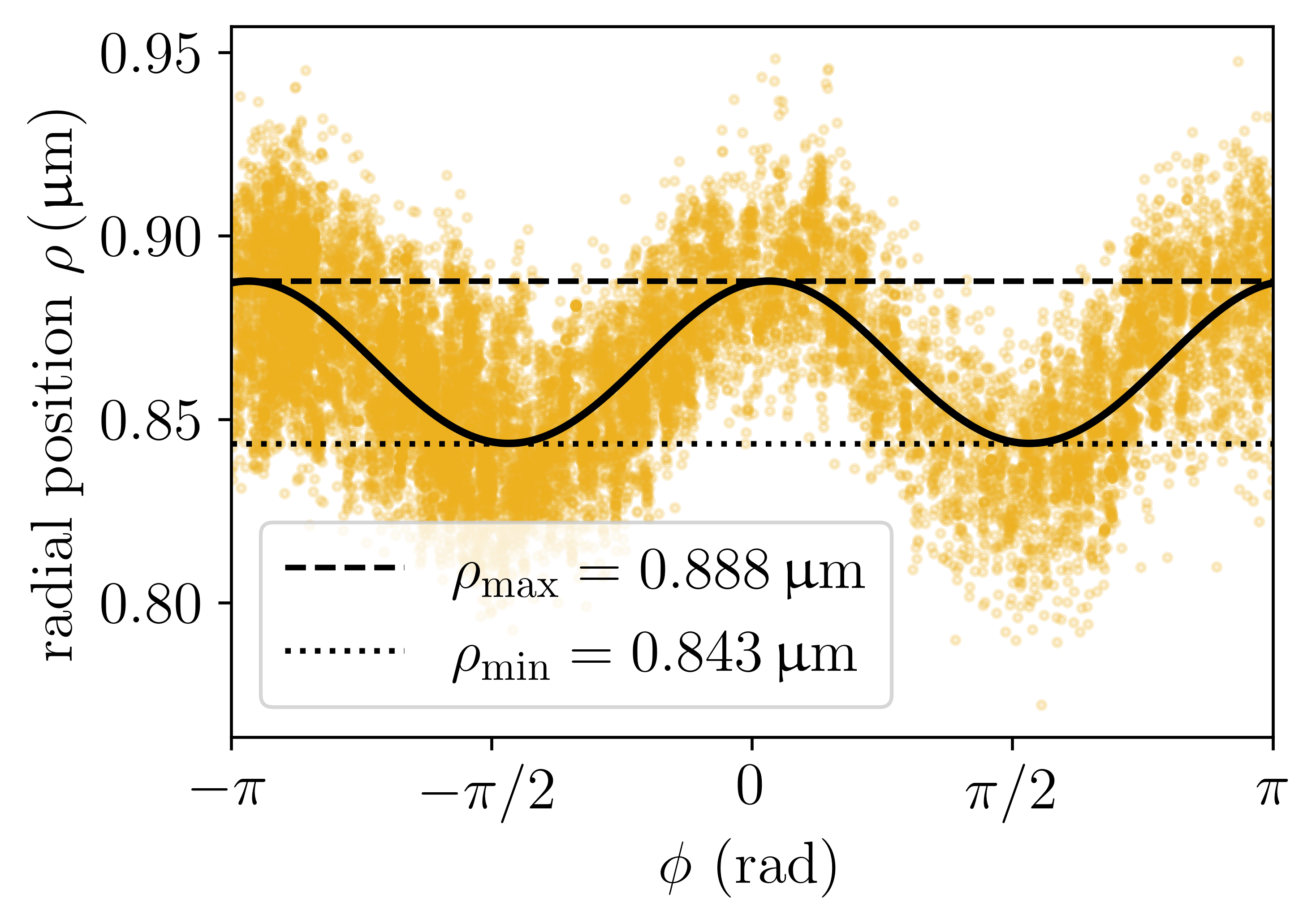}};
\node[
  anchor=north west,
  font=\scriptsize
] at ([xshift=0pt,yshift=11pt] img.north west) {(b)};
\end{tikzpicture}
\end{subfigure}

    \begin{subfigure}{0.64\linewidth}
\centering
\begin{tikzpicture}
\node[inner sep=0] (img)
  {\includegraphics[width=1\linewidth]{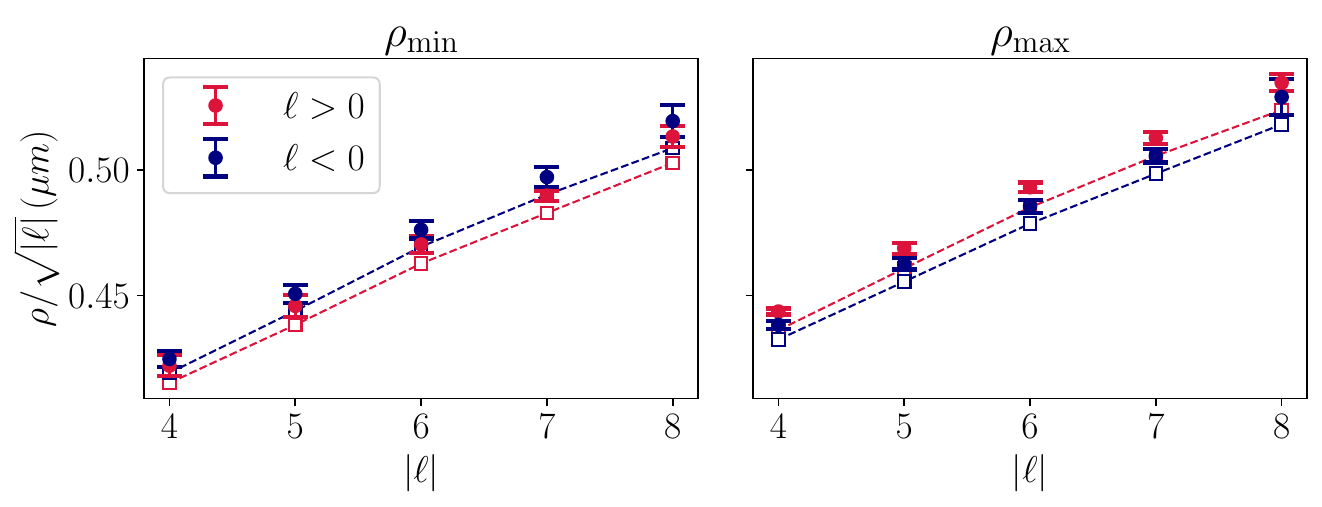}};
\node[
  anchor=north west,
  font=\scriptsize
] at ([xshift=15pt,yshift=3pt] img.north west) {(c)};
\end{tikzpicture}
\end{subfigure}

    \caption{(a) Detected position of control bead a in the $xy$-plane for all topological charges. 
    (b) Radial position for $\ell=4$ as a function of the azimuth angle $\phi$. The black line shows the result of an ellipse fit.
    (c) Average minimal and maximal radial equilibrium (circle symbols) across the control beads as a function of the topological charge $\ell$.  
    The square symbols represent the results obtained from a fit with the MDSA+ theory. 
    }
    \label{fig:rheq_beadA_ell4}
\end{figure}

First, we describe how the orbit parameters were extracted from the experimentally measured positions of the trapped control beads. 
In particular, we are interested in the minimum and maximum radial positions.
The particle positions were measured for four control beads, labeled a–d, and for Laguerre–Gaussian beams with topological charges $|\ell| = 4, 5, 6, 7, 8$.
To determine the orbit parameters, we fitted the measured trajectories on the $xy-$plane to an ellipse. We therefore transformed the measured coordinates into the reference frame of the ellipse by applying a rotation by an angle $\phi$ and a translation by $(x_c, y_c)$, according to
\begin{equation}
    \begin{pmatrix}
        x' \\ y'
    \end{pmatrix}
    = \begin{pmatrix}
        \cos \phi & \sin\phi
        \\ -\sin\phi & \cos\phi
    \end{pmatrix}
      \begin{pmatrix}
        x-x_c \\ y -y_c
    \end{pmatrix}
\end{equation}
and then fitted the data to the ellipse equation
\begin{equation}
    \frac{{x'}^2}{a^2} + \frac{{y'}^2}{b^2} = 1
\end{equation}
which allowed us to extract the minimum $\rho_\text{min}$ and maximum $\rho_\text{max}$ radial positions of the bead, as shown for bead a with $\ell = 4$ in Fig.~\ref{fig:rheq_beadA_ell4}.
For simplicity, we assumed in the following that the particle position precision is equal to the pixel size in each frame ($27.3\, \text{px} = 1\,\microns$), leading to an error of $\sigma_\text{pix} = 0.04\,\microns$.

The direction of astigmatism is not relevant for our analysis. Therefore, we set $\phi_\text{ast} = 0$ in all theoretical calculations.
As a consistency check, we verified that choosing $\phi_\text{ast}= \pi/4$ yields identical results.
Furthermore, we fixed the height of the objective's focal point with respect to the coverslip as $L = 1\,\microns$ in order to simplify the calculations. 
The variation of the radial equilibrium position with changing the axial displacement was found to be small compared to the experimental uncertainty of the measured positions. Therefore, $L$ was kept fixed for determining $A_\text{ast}.$

The astigmatism amplitude $A_{\mathrm{ast}}$ was determined by fitting the experimental values of the minimum and maximum equilibrium positions to the corresponding predictions of the MDSA+ theory. Since  astigmatism originates from the optical setup, it is assumed to be the same for all beads. 
However, the effective astigmatism may vary with the topological charge of the beam. 
Therefore, a separate fit was performed for each topological charge $\ell$, using the data from all four control beads. 
The fitting results for $A_\text{ast}$ are given in Table~\ref{tab:fitted_Aast} and the resulting values of $\rho_\text{max}$ and
$\rho_\text{min}$ are depicted in Fig.~\ref{fig:rheq_beadA_ell4} together with the average values extracted from the experiment and their standard deviation. 

\begin{table}[htbp]
    \centering
    \renewcommand{\arraystretch}{1.1}
    \setlength{\tabcolsep}{5pt}
    \begin{subtable}{0.3\linewidth}
    \begin{tabular}{c|c|c}
    $\ell$ & $A_\text{ast}$ & $\chi^2$
    \\ \hline 
    4 & $0.03 \pm 0.02$ & 1 \\
    5 & $0.03 \pm 0.02$ & 2 \\
    6 & $0.03 \pm 0.02$ & 2 \\
    7 & $0.03 \pm 0.01$ & 2 \\
    8 & $0.03 \pm 0.01$ & 6 \\
    \end{tabular}
    \end{subtable}
    \begin{subtable}{0.3\linewidth}
        \begin{tabular}{c|c|c}
    $\ell$ & $A_\text{ast}$ & $\chi^2$
    \\ \hline 
    -4 & $0.02 \pm 0.02$ & 1 \\
    -5 & $0.02 \pm 0.02$ & 2 \\
    -6 & $0.01 \pm 0.02$ & 2 \\
    -7 & $0.01 \pm 0.01$ & 2 \\
    -8 & $0.01 \pm 0.01$ & 7 \\
    \end{tabular}
    \end{subtable}
    \caption{Fitted astigmatism parameters $A_\text{ast}$ from the control beads for each topological charge. $\chi^2$ determines the quality of the fit.}
    \label{tab:fitted_Aast}
\end{table}

\section{Determining the Pasteur parameter for the composites}

In this section, we present fits of the MDSA+ theory to the measured orbital periods, using the filling fraction $f$ and the Pasteur parameter $\kappa$ as fitting parameters. 
Specifically, we minimized the weighted mean-squared deviation between the average experimental period, $\bar{T}$, and the theoretical period, $T_{\text{MDSA+}}$, across all topological charge for each composite. 
The difference is weighted by the standard deviation $\sigma_{\ell}$ of the measured periods. 
\begin{equation}
    \chi^2 = \sum_{\ell} \left(
    \frac{\overline{T}_\ell - T_{\text{MDSA+,} \ell}}{\sigma_\ell}
    \right)^2 
\end{equation}

We present two different fitting procedures in the following. 
The first method is the one presented in the main article, where we take a fixed value for the displacement of the objective's focal point from the glass slide $L = 0.87\,\microns$ (see Fig.~\ref{fig:parameters}) as determined from the experimental procedure, for all topological charges.
In the second method, we perform an additional fitting step, where we extract the average displacement from the control beads for negative and positive topological charges and use them for the fitting of the composites. 
The motivation for this second approach will be discussed later.

\subsection{Method 1 - Fixed axial displacement}

\begin{figure}
    \centering
    \includegraphics[height=7cm]{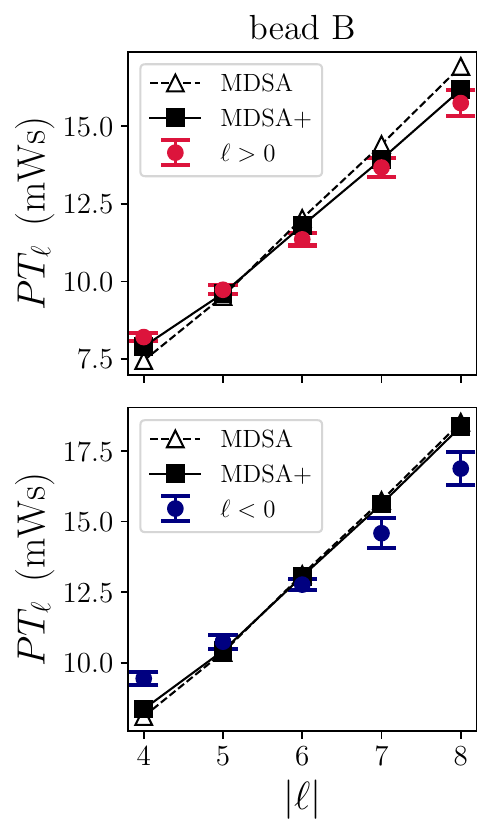}
    \includegraphics[height=7cm]{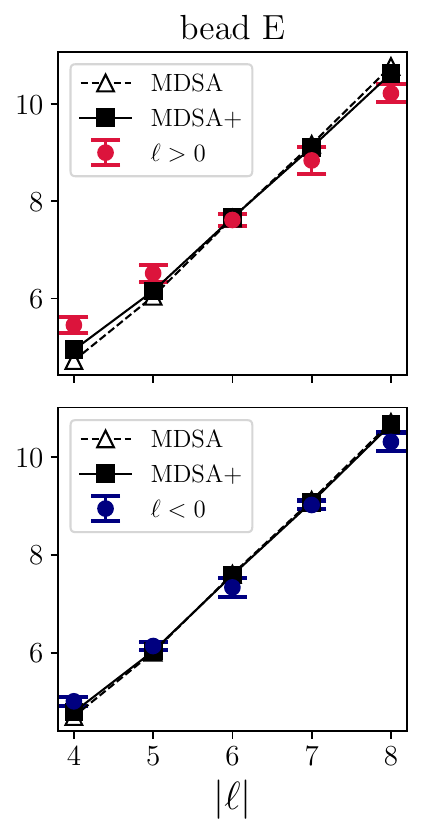}
    \includegraphics[height=7cm]{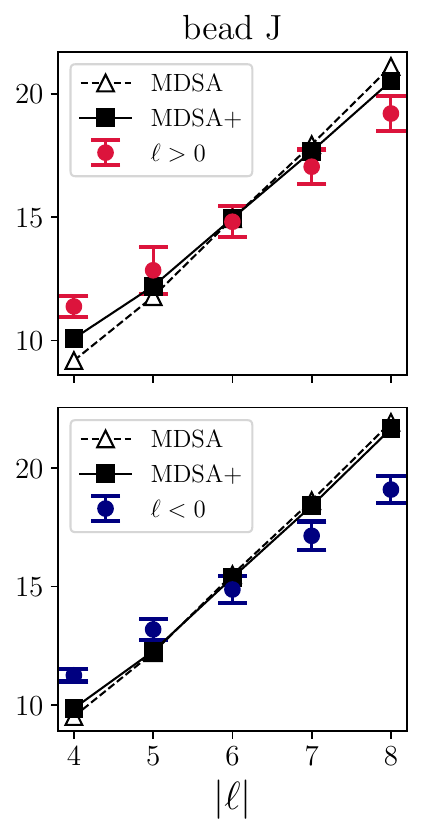}
    \includegraphics[height=7cm]{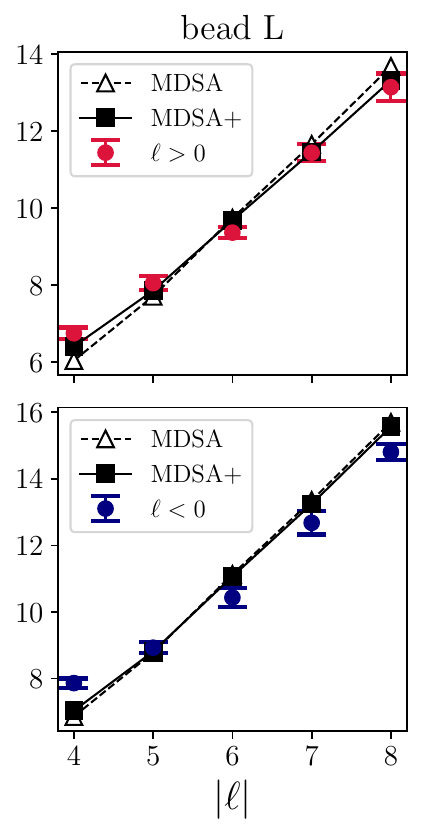}
    \caption{Experimentally measured orbital period for composites B, E, J and L. 
    The results are shown for positive (red circles) and negative (blue circles) topological charges. 
    Depicted are also the fitting results with the MDSA+ (squares) and MDSA (triangle) theory, where we used the nominal focal distance $L$. }
    \label{fig:fit_fixedL}
\end{figure}

The fitting results for the filling fraction and Pasteur parameter for a fixed displacement are shown in Table~\ref{tab:fit_MDSA+_MDSA_fixed_h_fixed_L} and depicted in Fig.~\ref{fig:fit_fixedL} for a few examples.
We present both results for the MDSA+ and MDSA theory. 
The results are qualitatively similar, but the MDSA+ theory fits the experimental data better, as can be seen by the smaller $\chi^2$ value across all beads. 

\begin{table}[h]
\centering
\renewcommand{\arraystretch}{1.1}
\setlength{\tabcolsep}{5pt}
\begin{subtable}[t]{0.45\textwidth}
\caption{MDSA+}
    \begin{tabular}{c|ccc}
        bead &  $f$ & $\kappa$ 
         & $\chi^2$
        \\ \hline
        A &
        $0.035 \pm 0.003$ &  
        $0.016 \pm 0.003$ &  
             34
        \\ \hline
        B &
        $0.093 \pm 0.009$ &  
        $0.015 \pm 0.001$ &  
             51
        \\ \hline
        C &
        $0.044 \pm 0.002$ &  
        $0.007\pm 0.002$ &  
             22
        \\ \hline
        D &
        $0.042 \pm 0.003$ &  
        $0.004 \pm 0.002$ &  
             32
        \\ \hline
        E &
        $0.156 \pm 0.001$ &  
        $0.002 \pm 0.002$ &  
             34
        \\ \hline
        F &
        $0.088 \pm 0.001$ &  
        $0.052 \pm 0.002$ &  
             79
        \\ \hline
        G &
        $0.021 \pm 0.003$ &  
        $0.006 \pm 0.002$ &  
             76
        \\ \hline
        H &
        $0.067 \pm 0.002$ &  
        $0.004 \pm 0.002$ &  
             72
        \\ \hline
        I &
        $0.040 \pm 0.003$ &  
        $0.009 \pm 0.002$ &  
             32
        \\ \hline
        J &
        $0.063 \pm 0.002$ &  
        $0.007 \pm 0.002$ &  
             72
        \\ \hline
        K &
        $0.002 \pm 0.001$ &  
        $0.016 \pm 0.006$ &  
            89
        \\ \hline
        L &
        $0.118 \pm 0.001$ &  
        $0.019 \pm 0.002$ &  
             66
        \\ \hline
        M &
        $0.061 \pm 0.002$ &  
        $0.016 \pm 0.002$ &  
             63
        \\ \hline
        N &
        $0.062 \pm 0.001$ &  
        $0.018 \pm 0.002$ &  
             16
    \end{tabular}
    \end{subtable}    
    \begin{subtable}[t]{0.45\textwidth}
    \caption{MDSA}
    \begin{tabular}{c|ccc}
        bead &  $f$ & $\kappa$ 
        &  $\chi^2$ 
        \\ \hline
        A &
        $0.021 \pm 0.004$ &  
        $0.012 \pm 0.002$ &  
         62  
        \\ \hline
        B &
        $0.088 \pm 0.001$ &  
        $0.009 \pm 0.001$ &  
         120  
        \\ \hline
        C &
        $0.035 \pm 0.002$ &  
        $0.003 \pm 0.002$ &  
         64  
        \\ \hline
        D &
        $0.034 \pm 0.003$ &  
        $-0.001 \pm 0.002$ &  
         53  
        \\ \hline
        E &
        $0.154 \pm 0.001$ &  
        $-0.001 \pm 0.002$ &  
         61  
        \\ \hline
        F &
        $0.084 \pm 0.001$ &  
        $0.047 \pm 0.002$ &  
         128 
        \\ \hline
        G &
        $0.009 \pm 0.004$ &  
        $0.002 \pm 0.002$ &  
         114  
        \\ \hline
        H &
        $0.063 \pm 0.002$ &  
        $0.001 \pm 0.002$ &  
         100  
        \\ \hline
        I &
        $0.033 \pm 0.003$ &  
        $0.007 \pm 0.002$ &  
         62  
        \\ \hline
        J &
        $0.058 \pm 0.002$ &  
        $0.004 \pm 0.002$ &  
         117  
        \\ \hline
        K &
        $0.000 \pm 0.002$ &  
        $0.014 \pm 0.003$ &  
         128  
        \\ \hline
        L &
        $0.114 \pm 0.001$ &  
        $0.016 \pm 0.002$ &  
         113 
        \\ \hline
        M &
        $0.053 \pm 0.002$ &  
        $0.010 \pm 0.002$ &  
         125  
        \\ \hline
        N &
        $0.056 \pm 0.001$ &  
        $0.013 \pm 0.002$ &  
         45  
    \end{tabular}
    \end{subtable}
\caption{Fitting results obtained from the fit with the MDSA+ and MDSA theory for fixed $L$.}
\label{tab:fit_MDSA+_MDSA_fixed_h_fixed_L}
\end{table}

We also tested the robustness of our method by performing a fit where we set $\kappa=0$ and instead used the shell thickness $h$ as fitting parameter. 
The fit was done for bead F, which showed the largest value for $\kappa$. 
We used the MDSA+ theory and found that the quality of the fit is much worse ($\chi^2 = 667$ compared to $79$) than for a non-zero Pasteur parameter (see Table~\ref{tab:fit_fixed_kappa_fixed_L} and Fig.~3(b) of the letter).

\begin{table}[h]
\centering
\renewcommand{\arraystretch}{1.1}
\setlength{\tabcolsep}{5pt}
\begin{tabular}{c|ccc}
bead &  $f$ & $h\, (\microns)$ 
&  $\chi^2$
\\ \hline
F &
$0.15 \pm 0.17$ &  
$0.04 \pm 0.06$ &  
  667
\end{tabular}
\caption{Fitting result for fixed chirality index $\kappa=0$ within the MDSA+ theory.}
\label{tab:fit_fixed_kappa_fixed_L}
\end{table}

The found Pasteur parameters show a bias towards positive values. 
This bias is already present in the control beads, and can also be seen from the plot of the dissymmetry factor shown in Fig.~3(a) of the letter. 
To account for this asymmetry, we perform a fit of the control beads using the same chiral core-shell model as for the composites. 
The results are given in Table~\ref{tab:control_beads_fixedL}. 
As can be seen, the values found for $\kappa$ are positive across all beads. 
We used the average of these values to correct the Pasteur parameters of the composites (see Fig.~3(c) of the letter). 

\begin{table}[h]
\centering
\renewcommand{\arraystretch}{1.1}
\setlength{\tabcolsep}{5pt}
\begin{subtable}{0.45\linewidth}
\caption{MDSA+}
\begin{tabular}{c|ccc}
bead &  $f$ & $\kappa$ 
 & $\chi^2$
\\ \hline
a &
$0.029 \pm 0.003$ &  
$0.014 \pm 0.002$ &  
48
\\ \hline
b &
$0.017 \pm 0.004$ &  
$0.004 \pm 0.002$ &  
15
\\ \hline
c &
$0.000 \pm 0.003$ &  
$0.013 \pm 0.002$ &  
45
\\ \hline
d &
$0.021 \pm 0.004$ &  
$0.020 \pm 0.002$ &  
27
\end{tabular}
\end{subtable}
\begin{subtable}{0.45\linewidth}
\caption{MDSA}
\begin{tabular}{c|ccc}
bead  &  $f$ & $\kappa$ 
 & $\chi^2$
\\ \hline
a &
$0.013 \pm 0.005$ &  
$0.007 \pm 0.002$ &  
64 
\\ \hline
b &
$0.007 \pm 0.006$ &  
$0.002 \pm 0.002$ &  
31 
\\ \hline
c &
$0.000 \pm 0.001$ &  
$0.011 \pm 0.002$ &  
87
\\ \hline
d &
$0.006 \pm 0.006$ &  
$0.017 \pm 0.002$ &  
41 
\end{tabular}
\end{subtable}
\caption{Fit of the control beads using the chiral core-shell model. }
\label{tab:control_beads_fixedL}
\end{table}

%\clearpage
%\newpage

\subsection{Method 2 - Axial displacement of the focal point determined from the control beads}\label{sec:method2}

One systematic effect that would explain at least part of the bias mentioned in the previous paragraph is the possibility that the optical paths for positive and negative topological charges differ. 
Indeed, the laser beam with negative topological charges undergoes a path which is slightly longer than in the case of positive ones, as discussed in connection with Fig.~1(a) of the letter. 
As a consequence, the laser beam corresponding to negative modes may be slightly converging at the back focal plane of the objective, resulting in a systematically smaller effective displacement length $L$ when compared to positive modes.

\begin{figure}[h]
    \centering
    \includegraphics[width=0.8\linewidth]{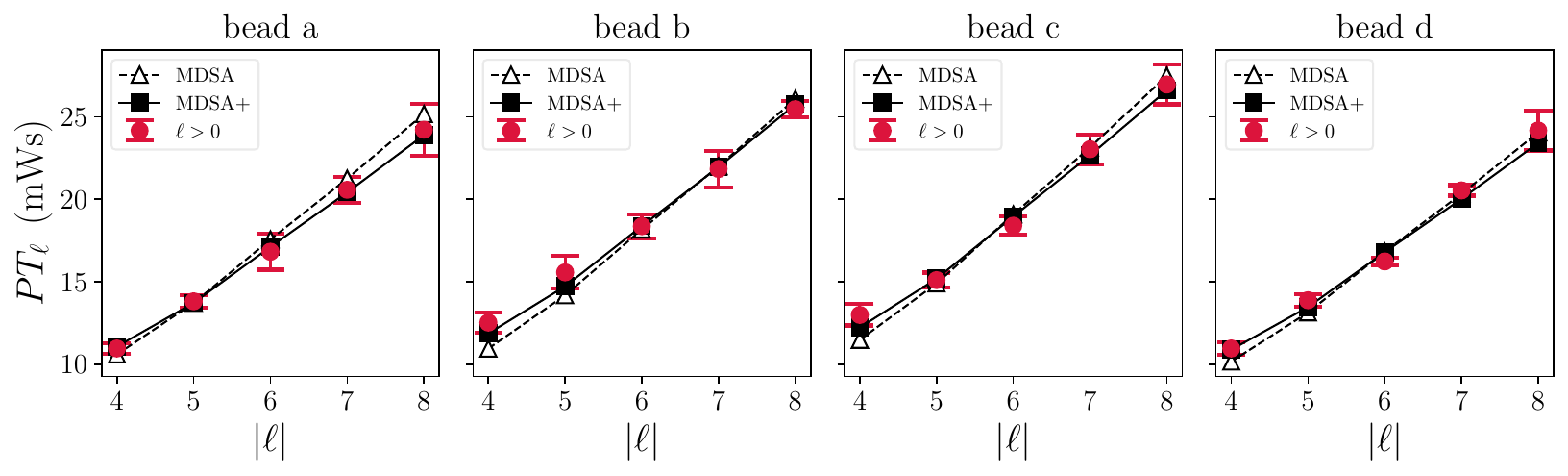}
    \includegraphics[width=0.8\linewidth]{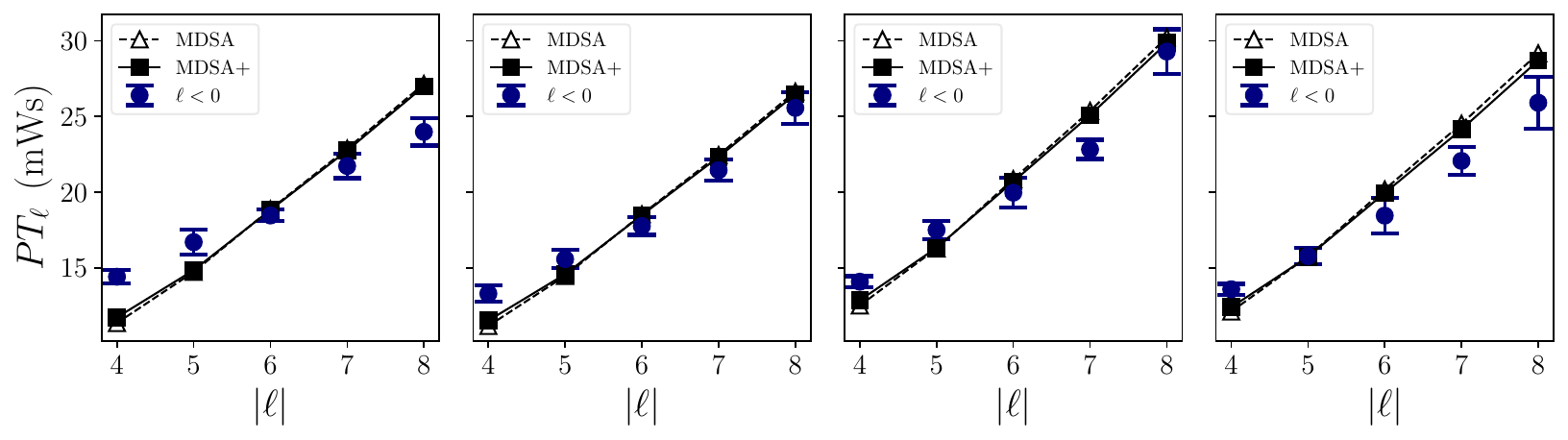}
    \caption{Experimentally measured orbital periods of the four control beads a-d for positive (red circles) and negative (blue circles) topological charges. 
    Additionally, the results of the fit with the MDSA+ (square) and MDSA (triangle) are shown. The displacement lengths $L_+$ and $L_-$ are used as fitting parameters for the positive and negative topological charges, respectively.}
    \label{fig:fit_control_beads}
\end{figure}

To investigate this effect, we used the measured orbital periods of the control beads to fit the displacement length $L$. 
The results obtained from the MDSA+ and MDSA theories are summarized in Table~\ref{tab:fitted_focaldistance_mdsaplus} and 
Table~\ref{tab:fitted_focaldistance_mdsa}, respectively, and depicted in Fig.~\ref{fig:fit_control_beads}. 
As can be seen, the fitted displacement is consistently smaller for negative topological charges across all control beads.

\begin{table}[h]
    \centering
    \renewcommand{\arraystretch}{1.1}
\setlength{\tabcolsep}{5pt}
\begin{subtable}[t]{0.45\textwidth}
\caption{MDSA+}
    \begin{tabular}{c||c|c||c|c}
    bead & $L_{+} (\microns)$ & $\chi_{+}^2$ & $L_{-} (\microns)$ & $\chi_{-}^2$
    \\ \hline
    a & $0.92 \pm 0.07$ & 0 & $0.61 \pm 0.02$ & 54 \\
    b & $0.71 \pm 0.03$ & 2 & $0.63 \pm 0.03$ & 17 \\
    c & $0.65 \pm 0.03$ & 3 & $0.50 \pm 0.01$ & 29 \\
    d & $1.02 \pm 0.06$ & 9 & $0.54 \pm 0.02$ & 21 \\
    \end{tabular}
    \label{tab:fitted_focaldistance_mdsaplus}
\end{subtable}
\begin{subtable}[t]{0.45\textwidth}
    \caption{MDSA}
     \begin{tabular}{c||c|c||c|c}
    bead & $L_+ (\microns)$ & $\chi^2$ & $L_- (\microns)$ & $\chi^2$
    \\ \hline
    a & $0.72 \pm 0.04$ & 3  & $0.59 \pm 0.02$ & 67 \\
    b & $0.65 \pm 0.03$ & 10 & $0.62 \pm 0.03$ & 23 \\
    c & $0.58 \pm 0.02$ & 7 & $0.49 \pm 0.01$ & 40 \\
    d & $0.84 \pm 0.04$ & 15 & $0.52 \pm 0.02$ & 30 \\
    \end{tabular}
    \label{tab:fitted_focaldistance_mdsa}
\end{subtable}
\caption{Fitted focal distance $L_{\pm}$ within the MDSA+ and MDSA theory for each control bead.}
\label{tab:fitted_focaldistance}
\end{table}

\begin{figure}[h]
    \centering
    \includegraphics[height=6.5cm]{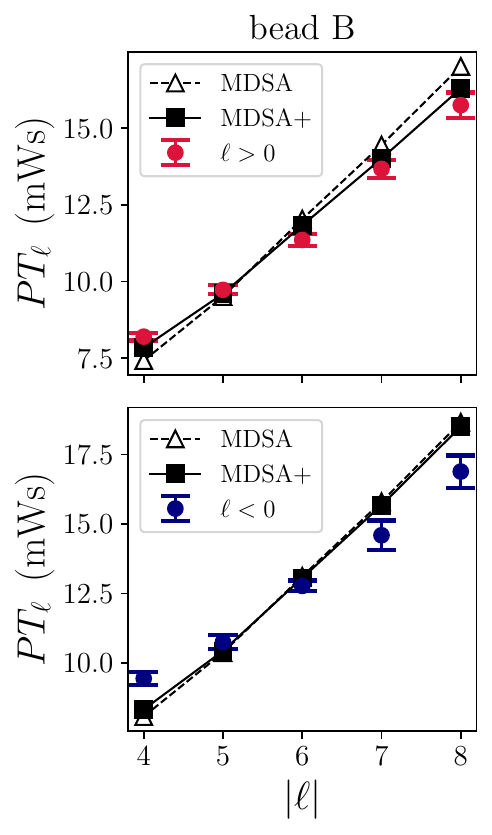}
    \includegraphics[height=6.5cm]{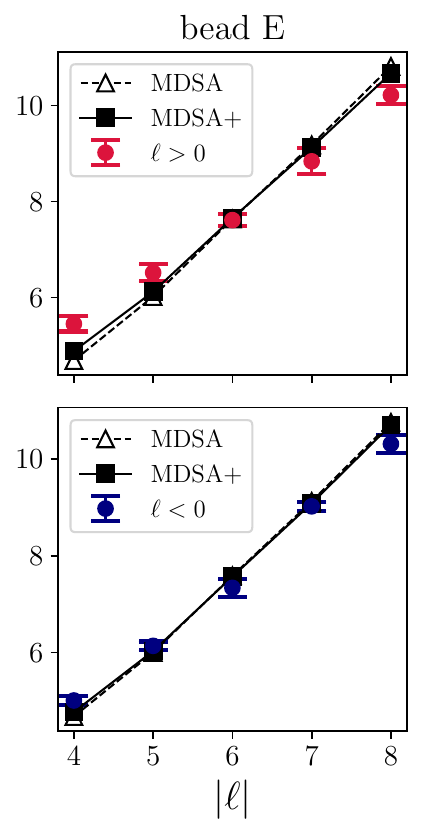}
    \includegraphics[height=6.5cm]{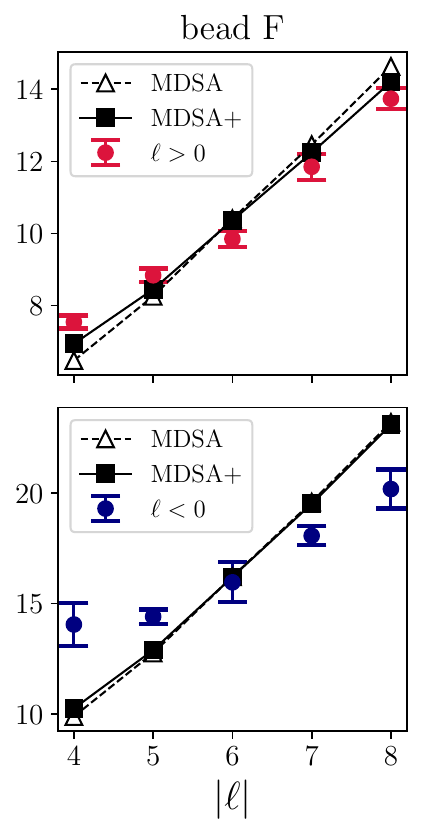}
    \includegraphics[height=6.5cm]{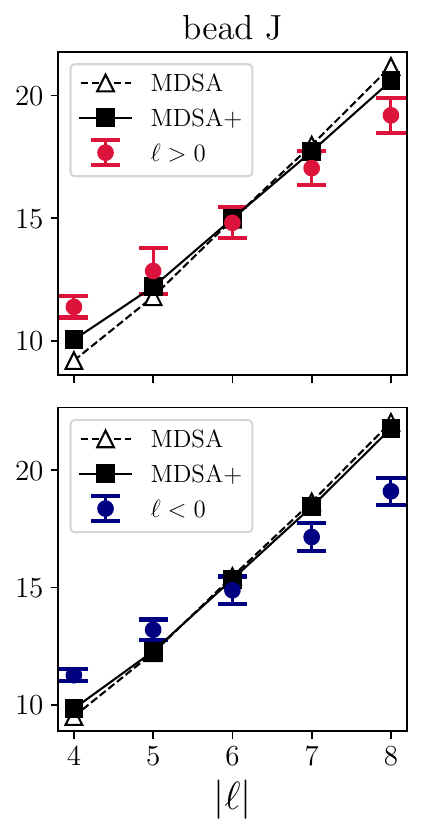}
    \includegraphics[height=6.5cm]{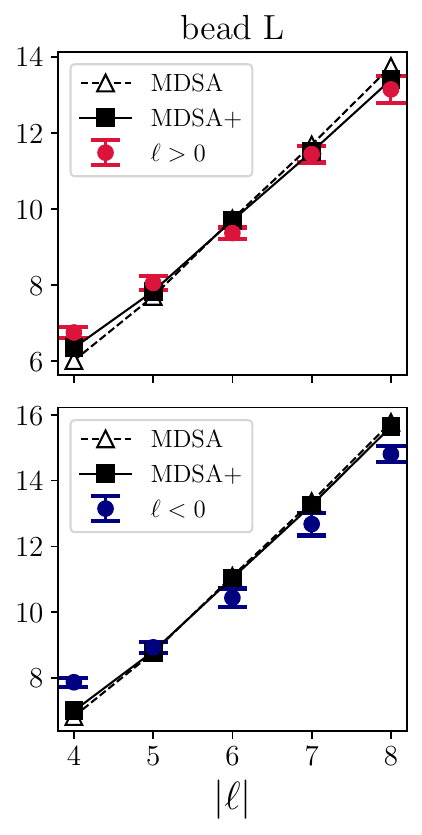}
    \caption{Same as in Fig.~\ref{fig:fit_fixedL} but using the values of $L_\pm$ obtained from fitting the periods of the control beads. 
    Also, the plot for bead F was added.}
    \label{fig:fit_averageL}
\end{figure}

\begin{figure}[h]
    \centering
    \includegraphics[width=0.4\linewidth]{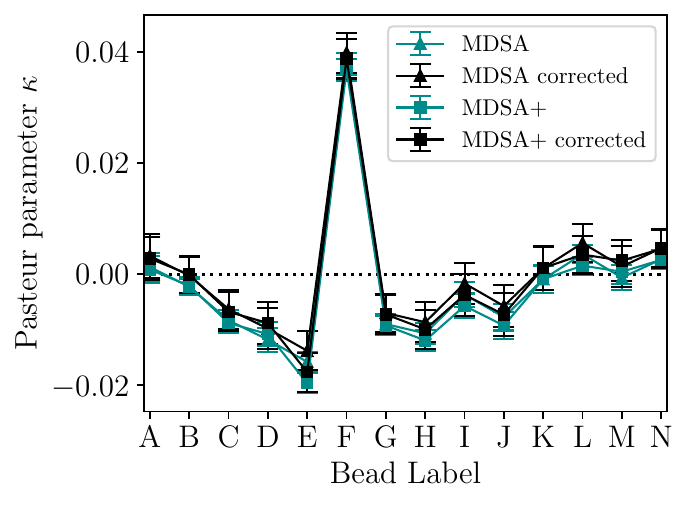}
    \caption{Pasteur parameter $\kappa$ for each bead found from the fit with the MDSA+ (teal circle) and MDSA theory (teal triangle) by using the average $L_\pm$ values. The black symbols represent the corrected values after subtraction of the offset value $\kappa = -0.002\pm 0.003$ determined from the control beads with the same method.}
    \label{fig:chirality_averageL}
\end{figure}

Using the MDSA theory, we obtained average values of $\bar{L}_+ = 0.70\,\microns$ and $\bar{L}_- = 0.56\,\microns$. 
Similarly, the MDSA+ theory yields $\bar{L}_+ = 0.82\,\microns$ and $\bar{L}_- = 0.57\,\microns$.
These average values were subsequently used as input parameters when fitting the composite particles.
The fitting results are summarized in Table~\ref{tab:fit_averageL} and depicted for some examples in Fig.~\ref{fig:fit_averageL}. 
The quality of the fit is comparable to the previous one, but the values of the filling fractions are slightly larger, and the chirality parameters are overall shifted to negative values, as can be seen from Fig.~\ref{fig:chirality_averageL}. 
In short, taking into account that $L_+>L_-$ explains part of the bias we observed from the previous method. 

\begin{table}[h]
\centering
\renewcommand{\arraystretch}{1.1}
\setlength{\tabcolsep}{5pt}
\begin{subtable}{0.45\linewidth}
\caption{MDSA+ }
\begin{tabular}{c|ccc}
bead &  $f$ & $\kappa$ 
 & $\chi^2$
\\ \hline
A &
$0.055 \pm 0.002$ &  
$0.001 \pm 0.002$ &  
 34 
\\ \hline
B &
$0.101 \pm 0.001$ &  
$-0.002 \pm 0.001$ &  
 58 
\\ \hline
C &
$0.062 \pm 0.002$ &  
$-0.009 \pm 0.002$ &  
 22 
\\ \hline
D &
$0.061 \pm 0.002$ &  
$-0.011 \pm 0.002$ &  
 32 
\\ \hline
E &
$0.166 \pm 0.001$ &  
$-0.020 \pm 0.002$ &  
 40 
\\ \hline
F &
$0.101 \pm 0.001$ &  
$0.037 \pm 0.002$ &  
85 
\\ \hline
G &
$0.045 \pm 0.002$ &  
$-0.009 \pm 0.002$ &  
72 
\\ \hline
H &
$0.081 \pm 0.001$ &  
$-0.012 \pm 0.002$ &  
75 
\\ \hline
I &
$0.059 \pm 0.002$ &  
$-0.006 \pm 0.002$ &  
32
\\ \hline
J &
$0.078 \pm 0.002$ &  
$-0.009 \pm 0.002$ &  
76
\\ \hline
K &
$0.030 \pm 0.004$ &  
$-0.001 \pm 0.002$ &  
74
\\ \hline
L &
$0.129 \pm 0.001$ &  
$0.002 \pm 0.002$ &  
74
\\ \hline
M &
$0.076 \pm 0.002$ &  
$0.000 \pm 0.002$ &  
67
\\ \hline
N &
$0.077 \pm 0.001$ &  
$0.003 \pm 0.002$ &  
18
\end{tabular}
\end{subtable}
\begin{subtable}{0.45\linewidth}
\caption{MDSA}
\begin{tabular}{c|ccc}
bead &  $f$ & $\kappa$ 
 & $\chi^2$
\\ \hline
A &
$0.052 \pm 0.003$ &  
$0.001 \pm 0.002$ &  
 60 
\\ \hline
B &
$0.105 \pm 0.001$ &  
$-0.002 \pm 0.001$ &  
 126 
\\ \hline
C &
$0.061 \pm 0.002$ &  
$-0.008 \pm 0.002$ &  
 64 
\\ \hline
D &
$0.060 \pm 0.002$ &  
$-0.012 \pm 0.002$ &  
 53 
\\ \hline
E &
$0.168 \pm 0.001$ &  
$-0.016 \pm 0.002$ &  
 68 
\\ \hline
F &
$0.102 \pm 0.001$ &  
$0.038 \pm 0.002$ &  
 133 
\\ \hline
G &
$0.046 \pm 0.002$ &  
$-0.009 \pm 0.002$ &  
 109 
\\ \hline
H &
$0.083 \pm 0.001$ &  
$-0.011 \pm 0.002$ &  
 102 
\\ \hline
I &
$0.060 \pm 0.002$ &  
$-0.004 \pm 0.002$ &  
 61 
\\ \hline
J &
$0.079 \pm 0.002$ &  
$-0.008 \pm 0.002$ &  
 121 
\\ \hline
K &
$0.030 \pm 0.004$ &  
$-0.001 \pm 0.002$ &  
 104 
\\ \hline
L &
$0.130 \pm 0.001$ &  
$0.004 \pm 0.002$ &  
 122 
\\ \hline
M &
$0.075 \pm 0.002$ &  
$-0.001 \pm 0.002$ &  
 127 
\\ \hline
N &
$0.077 \pm 0.001$ &  
$0.003 \pm 0.002$ &  
 46 
\end{tabular}
\end{subtable}
\caption{Results from a fit with the MDSA+ and MDSA theory with the average values for $L_{\pm}$ as determined from the control beads.}
\label{tab:fit_averageL}
\end{table}

To account for any remaining asymmetry, we again use the chiral core-shell model to fit the control beads. 
The results are shown in Table~\ref{tab:control_beads_averageL}. 
In both theories, we obtain an average offset of $\bar{\kappa} = -0.002 \pm 0.003$. 
The corrected Pasteur parameters of the composites are shown in Fig.~\ref{fig:chirality_averageL}. 
The resulting Pasteur parameters are consistent, within the uncertainties, with those obtained using the first method introduced earlier.

\begin{table}[h]
\centering
\setlength{\tabcolsep}{5pt}
\begin{subtable}{0.45\linewidth}
\caption{MDSA+}
\begin{tabular}{c|ccc}
bead &  $f$ & $\kappa$ 
 & $\chi^2$
\\ \hline
a &
$0.051\pm 0.002$ &  
$0.000 \pm 0.003$ &  
47
\\ \hline
b &
$0.044 \pm 0.002$ &  
$-0.011 \pm 0.002$ &  
13
\\ \hline
c &
$0.027 \pm 0.004$ &  
$-0.003 \pm 0.002$ &  
25
\\ \hline
d &
$0.047 \pm 0.002$ &  
$0.005 \pm 0.002$ &  
26
\end{tabular}
\end{subtable}
\begin{subtable}{0.45\linewidth}
\caption{MDSA}
\begin{tabular}{c|ccc}
bead &  $f$ & $\kappa$ 
 & $\chi^2_\text{MDSA}$
\\ \hline
a &
$0.048 \pm 0.002$ &  
$-0.003 \pm 0.002$ &  
62 
\\ \hline
b &
$0.045 \pm 0.002$ &  
$-0.009 \pm 0.002$ &  
28 
\\ \hline
c &
$0.025 \pm 0.004$ &  
$-0.002 \pm 0.002$ &  
42
\\ \hline
d &
$0.045 \pm 0.002$ &  
$0.006 \pm 0.002$ &  
38 
\end{tabular}
\end{subtable}
\caption{ Fit of the control beads using the chiral core-shell model.  We consider the average values $L_+=0.82\,\microns$ and $L_-= 0.57\,\microns$ for the MDSA+ calculation; and    $L_+=0.70\,\microns$ and $L_-= 0.56\,\microns$ for the MDSA one.}
\label{tab:control_beads_averageL}
\end{table}

\clearpage
\newpage
\bibliography{references}